\documentstyle[12pt]{article}

 \catcode`\@=11
\ifcase\@ptsize
 \font\tenmsa=msam10
 \font\sevenmsa=msam7
 \font\fivemsa=msam5
 \font\tenmsb=msbm10
 \font\sevenmsb=msbm7
 \font\fivemsb=msbm5
\or
 \font\tenmsa=msam10 scaled \magstephalf
 \font\sevenmsa=msam8
 \font\fivemsa=msam6
 \font\tenmsb=msbm10 scaled \magstephalf
 \font\sevenmsb=msbm8
 \font\fivemsb=msbm6
\or
 \font\tenmsa=msam10 scaled \magstep1
 \font\sevenmsa=msam8
 \font\fivemsa=msam6
 \font\tenmsb=msbm10 scaled \magstep1
 \font\sevenmsb=msbm8
 \font\fivemsb=msbm6
\fi

\newfam\msafam
\newfam\msbfam
\textfont\msafam=\tenmsa  \scriptfont\msafam=\sevenmsa
  \scriptscriptfont\msafam=\fivemsa
\textfont\msbfam=\tenmsb  \scriptfont\msbfam=\sevenmsb
  \scriptscriptfont\msbfam=\fivemsb

\def\hexnumber@#1{\ifnum#1<10 \number#1\else
 \ifnum#1=10 A\else\ifnum#1=11 B\else\ifnum#1=12 C\else
 \ifnum#1=13 D\else\ifnum#1=14 E\else\ifnum#1=15 F\fi\fi\fi\fi\fi\fi\fi}

\def\msa@{\hexnumber@\msafam}
\def\msb@{\hexnumber@\msbfam}

\mathchardef\hbar="0\msb@7E
\def\checkmark{\mathhexbox\msa@58 }

\def\Bbb{\ifmmode\let\next\Bbb@\else
 \def\next{\errmessage{Use \string\Bbb\space only in math mode}}\fi\next}
\def\Bbb@#1{{\Bbb@@{#1}}}
\def\Bbb@@#1{\fam\msbfam#1}

\if@twoside
\else
\fi
\ifcase \@ptsize
\or % mods for 11 pt
\or % mods for 12 pt
\fi

\def\@citex[#1]#2{%
\if@filesw \immediate \write \@auxout {\string \citation {#2}}\fi
\@tempcntb\m@ne \let\@h@ld\relax \def\@citea{}%
\@cite{%
  \@for \@citeb:=#2\do {%
    \@ifundefined {b@\@citeb}%
      {\@h@ld\@citea\@tempcntb\m@ne{\bf ?}%
      \@warning {Citation `\@citeb ' on page \thepage \space undefined}}%
      {\@tempcnta\@tempcntb \advance\@tempcnta\@ne%
      \@tempcntb\number\csname b@\@citeb \endcsname \relax%
      \ifnum\@tempcnta=\@tempcntb %   Number follows previous--hold on to it
        \ifx\@h@ld\relax%
          \edef \@h@ld{\@citea\csname b@\@citeb\endcsname}%
        \else%
          \edef\@h@ld{\ifmmode{-}\else--\fi\csname b@\@citeb\endcsname}%
        \fi%
      \else%   %  non-successor--dump what's held and do this one
        \@h@ld\@citea\csname b@\@citeb \endcsname%
        \let\@h@ld\relax%
      \fi}%
    \def\@citea{,\penalty\@highpenalty\,}%
  }\@h@ld
}{#1}}
\@addtoreset{equation}{section}
\def\section{\@startsection {section}{1}{\z@}{-3.5ex plus -1ex minus
 -.2ex}{2.3ex plus .2ex}{\large\bf\centering}}
\def\subsection{\@startsection{subsection}{2}{\z@}{-3.25ex plus -1ex minus
 -.2ex}{1.5ex plus .2ex}{\sc}}
\def\topspace{\vphantom{\vrule height 3ex depth 0pt}}
\def\bottomspace{\vphantom{\vrule height 0pt depth 2ex}}
\gdef\@publabel{\hfil}
\gdef\@pubdate{\null}
\gdef\@pubnumber{\null}
\gdef\@author{\null}
\gdef\@title{\null}
\gdef\@abstract{\null}
\long\def\pubdate#1{\gdef\@pubdate{#1}}
\long\def\pubnumber#1{\gdef\@pubnumber{#1}}
\long\def\publabel#1{\gdef\@publabel{#1}}
\long\def\author#1{\gdef\@author{#1}}
\long\def\title#1{\gdef\@title{#1}}
\long\def\abstract#1{\gdef\@abstract{#1}}
\def\titlerelax{
\let\maketitle\relax
\let\settitleparameters\relax
\let\consolidatetitle\relax
\let\inittitlepage\relax
\let\finishtitlepage\relax
\let\titlepagecontents\relax
\let\multithanks\relax
\let\titlebaselines\relax
\let\@makepub\relax
\let\@maketitle\relax
\let\@makeauthor\relax
\let\@makeabstract\relax
\let\@maketitlenote\relax
\let\thanks\relax
\let\titlerelax\relax}
\def\titleclean
{\gdef\@titlenote{}
\gdef\@abstract{}
\gdef\@author{}
\gdef\@title{}
\gdef\@pubdate{}\gdef\@pubnumber{}\gdef\@publabel{}
\gdef\@dpublabel{}
}
\def\@makepub{\vbox to \z@{\hbox to \textwidth{\hfill
\@publabel \hfill
\llap{\parbox[t]{0.25\textwidth}{\raggedleft\@pubnumber}}}%
\vss}}
\def\@maketitle{\vskip 60pt \begin{center}
 {\LARGE \@title \par}
 \end{center}}
\def\@makeauthor{{%
\def\and{\smallskip {\normalsize \rm and\smallskip }}
\def\And{\medskip {\normalsize \rm and\\}\medskip}
\long\def\address##1{{\def\and{\\and\\}\medskip
				{\small \it \\##1\\}
}}
{\centering
 \vskip 3em
 \large \lineskip .75em
 \@author}
 \par}}
\def\@makedate{\vskip 1.5em
 {\raggedright \small \noindent\@pubdate \par}}
\def\@makeabstract{\vskip 1.5em
{\small
\begin{center}
{\bf ABSTRACT\vspace{-.5em}\vspace{0pt}}
\end{center}
\quotation \@abstract \endquotation}}
\def\maketitle{\titlepage
\let\footnotesize\small \setcounter{page}{0}
\@makepub
\vfil
\@maketitle
\@makeauthor
\vfil
\@makeabstract
\@thanks
\vfil
\@makedate
\if@restonecol\twocolumn \else \eject \fi
\titlerelax \titleclean
\setcounter{footnote}{0}
}

\catcode`\@=12

\begin{document}
\bibliographystyle{npb}

%\input macros
%a

%b
\let\b=\beta
\def\blank#1{}

%c
\def\cdd{{\cdot}}
\def\cev#1{\langle #1 \vert}
\def\cH{{\cal H}}
\def\comm#1#2{\bigl [ #1 , #2 \bigl ] }
\def\compact{ reductive}
\def\cont{\nonumber\\*&&\mbox{}}
\def\cO{{\cal O}}
\def\cul #1,#2,#3,#4,#5,#6.{\left\{ \matrix{#1&#2&#3\cr #4&#5&#6} \right\}}

%d
\def\dz{Dz}
\def\dz{\hbox{$d\kern-1.1ex{\raise 3.5pt\hbox{$-$}}\!\!z$}}
\def\dz{ \frac{d\!z}{2\pi i}}
%e
\def\en{\end{equation}}
\def\enn{\end{eqnarray}}
\def\eq{\begin{equation}}
\def\eqq{\begin{eqnarray}}

%f

%g

%h
\def\half#1{\frac {#1}{2}}

%i
\def\ip#1#2{\langle #1,#2\rangle}

%j

%k
\def\k{k}

%l

%m
\def\Mf#1{{M{}^{{}_{#1}}}}
\def\mno{{\textstyle {\circ\atop\circ}}}
\def\mod#1{\vert #1 \vert}

%n
\def\Nf#1{{N{}^{{}_{#1}}}}
\def\ni{\noindent}
\def\no{{\textstyle {\times\atop\times}}}
\def\no:#1:{\mno#1\mno}
\def\nox{{\scriptstyle{\times \atop \times}}}

%o

%p
\let\p=\phi
\def\posdef{ positive-definite}
\def\posdefness{ positive-definiteness}
%q
\def\Qf#1{{Q{}^{{}_{#1}}}}
\def\Qstar{\mathop{\no:QQ:}\nolimits}

%r
\def\reductive#1#2{#1}

%s

%t
\def\tr{\mathop{\rm tr}\nolimits}
\def\Tr{\mathop{\rm Tr}\nolimits}

%u

%v

%r

%s

%t

%u

%v
\def\vec#1{\vert #1 \rangle}
\def\vac{\vec 0}

%w
\def\wan{$\WA_n$ }
\def\Wb{\bar W}
\def\Wf#1{{W{}^{{}_{#1}}}}
\def\wbn{$\WB_n$ }
\def\WA{\mathop{\it WA}\nolimits}
\def\WB{\mathop{\it WB}\nolimits}
\def\WBC{\mathop{\it WBC}\nolimits}
\def\WD{\mathop{\it WD}\nolimits}
\def\WG{\mathop{\it WG}\nolimits}

%x

%y

%z
\def\zz#1{(z-z')^{#1}}

\pubnumber{%{\bf DRAFT} \\
EFI 91-63\\DTP-91-63}
\pubdate{Nov. 1991}

\title{ On  the classification of quantum W-algebras}

\author{
P. BOWCOCK\thanks{ Email \tt
BOWCOCK@EDU.UCHICAGO.CONTROL}
\thanks{Supported by U.S. DOE grant  DEFG02-90-ER-40560 and NSF grant
PHY900036
}\address{Enrico Fermi Institute,
University of Chicago,
Chicago, IL 60637, U.S.A.}
\And
G. M. T WATTS\thanks{ Email \tt G.M.T.WATTS@UK.AC.DURHAM}\address{
Department of Mathematical Sciences, University of Durham, South Road,
Durham, DH1 3LE, U.K.}
}

\abstract{
In this paper we consider the structure of general quantum W-algebras.
We introduce the notions  of deformability,
positive-definiteness, and reductivity of a W-algebra.
We show that one can associate a reductive finite Lie algebra to each reductive
W-algebra.
The finite Lie
algebra is also endowed with a preferred $sl(2)$ subalgebra, which
gives the conformal weights of the W-algebra. We extend this to cover
W-algebras containing both bosonic and fermionic fields, and
illustrate our ideas with the Poisson bracket algebras of generalised
Drinfeld-Sokolov Hamiltonian systems. We then discuss the
possibilities of classifying deformable W-algebras which
fall outside this class in the context of automorphisms of Lie algebras.
In conclusion we list the
cases in which the W-algebra has no weight one fields, and further,
those in which it has only one weight two field.}

\maketitle

\section{Introduction}

In the last few years remarkable progress has been made in
the understanding of two-dimensional field theories that are
conformally
invariant. A key to completing this program is the classification of
extended conformal algebras, or W-algebras.
The first examples of W-algebras were the conformal algebra, various
superconformal algebras and Kac-Moody algebras \cite{GOli1}. Later it
was realised that
a wider variety of algebras could be constructed from GKO coset
theories \cite{GKOl,BBSS,Watt3,BGod1}.
Interest then moved to Toda theories, which provided many examples of
W-algebras \cite{GERV}; later these were re-incorporated into the
Drinfeld-Sokolov scheme of Hamiltonian reductions
\cite{BFFOW1,DSok,BOog1,ASha}.
Most recently it has been shown that the generalised Drinfeld-Sokolov
reduction of WZW models can be used to produce a new class of algebras.
In this construction one gauges a WZW model associated with a Lie algebra $g$
by the currents associated with some nilpotent subalgebra. For bosonic
W-algebras with fields of integer conformal weight, the nilpotent subalgebra
can be labelled by an (integral)
$su(1,1)$ embedding in $g$ \cite{Int}.
This latter method has the advantage that the properties of the algebra
obtained are easily related to the finite Lie algebraic ingredients of
the construction, while the corresponding relationship in the GKO case
is much more mysterious for the present.

However illuminating the examples cited above may be, we cannot hope to
obtain a classification scheme for W-algebras if we tie ourselves to
any one construction. It is this that motivates us to adopt a more
general standpoint in this paper. The study of W-algebras is
hampered by their infinite-dimensional nature. Worse still, their
commutation relations are generally non-linear in the generating
fields. Some progress has been made by looking for examples
using algebraic computing techniques
 \cite{KWat1,BFKNRV1}, %,CKwa1},
but the calculations involved are complex, and so the searches are restricted
to examples containing two or three fields of low conformal weight.

In this paper we shall restrict our attention to `deformable' algebras.
By deformable we mean that the algebra satisfies the Jacobi identity
for a continuous range of values of the central charge $c$. In general the
structure constants of the algebra are allowed to be functions of $c$.
The algebras excluded by this restriction are a very complicated set of
objects, which in principle include, for example, a large number of
W-algebras which can be constructed from lattices
using vertex operators \cite{DGMo,FLM}. There is some
hope that these algebras are
extensions of deformable algebras, occurring when a generically non-integer
weighted primary field becomes integer weighted for particular values of $c$.

The main result of our paper is contained in section 2, where
we demonstrate the existence of a finite subalgebra associated with each
classical W-algebra.
Perhaps of more importance is that we can extend this result to the quantum
case
if we demand that the quantum algebra
have a `good classical limit'. After discussing what precisely this means,
we demonstrate the existence of a similar finite subalgebra
in the limit that the central charge $c\to \infty$.
This provides us with an easily computable characteristic
for W-algebras.
A special role is  played by an $su(1,1)$ subalgebra of this
finite algebra, which corresponds to the modes $L_1,L_0, L_{-1}$ of the
Virasoro algebra which generate M\"obius transformations.

The characteristic we have derived is a finite dimensional Lie algebra, and
an $su(1,1)$ embedding. This is precisely the data used
in the generalised Drinfeld-Sokolov reduction method. In sections 3 and 4
we clarify this connection.
After a  review of the generalised Drinfeld-Sokolov
construction of $W$-algebras, we calculate the structure constants of the
$W$-algebra obtained by this method up to linear order in the fields, and
use this to demonstrate that in this case the finite
subalgebra is simply the Lie algebra $g$ associated with the WZW model that we
are reducing and that, further, the $su(1,1)$ embedding is the same as
that used
in the construction. This provides us with a proof of the existence of the
classical W-algebras associated with each finite Lie algebra $g$, and
$su(1,1)$   embedding.

Armed with these results we discuss certain features of the $W$-algebras that
are constructed in this way in section 5.
In particular we give a complete list of such
algebras which have no Kac-Moody components, and those with only one spin-$2$
field. We also comment on the possible use of automorphisms of the finite
Lie algebra to generate homomorphisms of the W-algebra. The resulting
algebras will be deformable, but will not have a good classical limit.

Finally, we conclude with some comments on the relevance of our approach to
the classification of deformable $W$-algebras.

\section{Finite algebras from $W$-algebras}

If we want to classify extended conformal symmetries, or $W$-algebras,
we should like to attribute to them
some easily computable characteristics which
specify the algebra completely. In this section we shall construct a
finite Lie algebra associated with classical $W$-algebras and their
quantum counterparts. Although we do not prove that this  specifies
the $W$-algebra,
it does reveal something of its structure, and may ultimately form
part of some classification scheme.
To start with we shall consider the relationship between a general
quantum $W$-algebra and its classical counterpart. After discussing
the `vacuum preserving algebra' (vpa) for both types of algebra, we
show that in the classical case this contains a finite subalgebra if
we define a `linearised' Poisson bracket. We then extend this
result to the quantum case by showing that the corresponding finite
algebra decouples in the limit that the central charge goes to infinity.

Let us begin by discussing the relationship between quantum W-algebras and
their classical counterparts.
A quantum W-algebra comprises a set of modes $W^a_m$ of some simple
fields $W^a(z)$, a notion of
normal ordering, and a Lie Bracket.
W-algebras are usually presented in the form of an operator product
expansion, which we
may represent schematically as
\eqq
W^a(z)W^b(z') &=& g^{ab} \zz{-\Delta^a - \Delta^b} \nonumber\\
&+& \sum_c
f^{ab}_{(1)}{}_c\zz{\Delta^c - \Delta^b - \Delta^a} [W^c(z') +
g^{ab}_c(z-z') \partial W^c(z') +
\ldots ] \label{eq.stqu} \\
&+& \sum_{c,d} f_{(2)}^{ab}{}_{cd}(z-z')^{\Delta^c + \Delta^d -
\Delta^a - \Delta^b}  [ \mno W^c(z')W^d(z')\mno + \ldots ] +
\ldots \; |z| > |z'| \nonumber
\enn
Here we have arranged the right hand side according to degree in $W$.
Since we
are interested in conformal field theories, we assume that the
algebra contains the Virasoro algebra
\eq
L(z) L(z') = {c\over 2}\zz{-4} + 2L(z')\zz{-2}+\partial L(z')\zz{-1}
+ O(1) \;,
\label{vir}
\en
as a subalgebra. We also assume that
the algebra is generated by a finite number of primary fields $W^a(z)$
which obey
\eq
L(z)W^a(z') = \Delta^a W^a(z')\zz{-2}+\partial W^a(z')\zz{-1}
+ O(1)
\;,
\label{eq.prim}
\en
where $\Delta^a$ is the weight of $W^a$.
The commutation relations of the modes $W^a_m$ which are given by
$W^a(z) = \sum W^a_mz^{-\Delta_a-m}$ can be deduced in the standard
manner from the operator product expansion by a double contour
integral.
These take the form
\eqq
{}~[W^a_m,W^b_n] &=& g^{ab}P(\Delta_a,\Delta_b,0,m,n) +
f^{ab}_{(1)c}P(\Delta_a,\Delta_b,\Delta_c,m,n)W^c_{m+n}  \nonumber\\
&&+ f^{ab}_{(2)cd}P(\Delta_a,\Delta_b,\Delta_c+\Delta_d,m,n)
\mno W^cW^d\mno_{m+n} + \ldots,
\label{eq.stqumodes}
\enn
where $P$ is some known polynomial.
In terms of modes (\ref {eq.prim}) becomes
\eq
{}~[L_m,W^a_n]= [(\Delta_a-1)m-n]W^a_{m+n}
\;.
\label{eq.pr}
\en
Any field that obeys (\ref {eq.pr})
for all $m$ is called {\it primary}, and any field that obeys it for $m=-1,0,1$
is called {\it quasi-primary}.
The modes of a quasi-primary field
$ \cO^i_m$ form an
indecomposable representation of the $su(1,1)$ algebra generated by
$L_{-1},L_0,L_1$.
One can use the representation theory of $su(1,1)$ to show that the
quasi-primary fields and their derivatives span the space of fields
in the algebra, and that the polynomials $P$
are related to Clebsch-Gordan coefficients.

We define the hermitian conjugate by
\eq
(W^a_m)^\dagger=W^a_{-m},
\label{eq.herm}
\en
and this induces a natural inner product on the states of the quantum theory.
The requirement that this inner product be positive definite and the
representation theory of the Virasoro algebra requires that the central charge
$c>0$ and that the fields all have positive definite weight.
This then implies
that the metric $g^{ab}$ is only non-vanishing of fields of equal weight, that
it is positive definite, and that a basis of fields satisfying (\ref{eq.herm})
can be chosen for which the metric is diagonal.
We call such a W-algebra {\it positive-definite}.
With this choice of basis,
the algebra (\ref{eq.stqumodes}) takes the form
\eqq
 \comm{W^a_m}{W^b_n} &=& (c/(2\Delta_a -1)!\Delta_a)
\delta^{ab}m(m^2-1)\ldots(m^2 - (\Delta_a-1)^2)\delta_{m+n,0}
\label{eq.stqumodes2}
\\
&& \kern -2truecm +
f^{ab}_{(1)c}P(\Delta_a,\Delta_b,\Delta_c,m,n)W^c_{m+n}
%\nonumber\\ &&
+ f^{ab}_{(2)cd}P(\Delta_a,\Delta_b,\Delta_c+\Delta_d,m,n)
\mno W^cW^d\mno_{m+n} + \ldots
\nonumber
\enn
where $f$ are constants, and we have used that
$P(\Delta_a,\Delta_a,0,m,n)=m(m^2-1)\cdots(m^2-(\Delta_a-1)^2)\delta_{m+n,0}$ .
We do not as yet require that the algebra be
defined for more than one value of $c$. Examples of algebras which are not
positive-definite include Kac-Moody algebras based on non-compact groups,
and algebras including `ghost' fields with strange statistics, such as the
bosonic $N=2$ superalgebra recently considered in \cite{Ber1}.

We should remark on the definition of normal ordering $\mno\mno$ which
we use here. In meromorphic conformal field theory, we assign a field
uniquely to each state by
\eq
\phi(z) \leftrightarrow \phi(0)\vac = \vec\phi
= \phi_{-\Delta_\phi}\vac
\;.
\en
We can define the normal ordered field $\nox \phi \phi'\nox$ by
\eq
\nox\phi\phi'\nox \leftrightarrow
\phi_{-\Delta_\phi}\phi'_{-\Delta_{\phi'}}\vac
\;.
\en
However this is not the only possible normal ordering.
Following Nahm,  \cite{n:talk,BFKNRV1}, we have
introduced the normal ordering $\mno\mno$, by
\eq
\mno \phi\phi'\mno = {\cal P} \nox \phi\phi' \nox
\;,
\en
where $\cal P$ is the projector onto $su(1,1)$ highest weight fields
so that the resulting composite fields are quasi-primary. Further
ambiguities arise when we try to normal order more than two fields,
since this product is not associative.
One can, for example, decide to order the
fields by conformal weight and index $a$, and then always nest the
normal orderings from the left. However, there are many choices of
basis. We shall call a particular choice of  basis a {\it presentation} of the
W-algebra as a commutator algebra. The underlying structure, which is
that of a meromorphic conformal field theory,
is the same for each presentation, but the
structure constants will be different. The point that we should like to stress
is that the classical limit of different presentations are identical. This
is a consequence of the observation that the difference in two orderings
can be written as a commutator, and thus must be an $O(\hbar)$ term which
vanishes in the classical limit.(In fact, the projection operator $\cal P$
does not simply amount to a reordering, but the difference between the
two normal orderings can be seen to be the Virasoro descendents of commutators,
so that the result is true in this case too.)

Let us now consider classical $W$-algebras. This is a Poisson
bracket algebra of fields $W^a(x)$ of one variable which closes on
(differential) polynomials and central terms. We can represent the
Poisson bracket schematically as
\eqq
&& \kern-1truecm\{ W^a(x), W^b(y)\} =
g^{ab}\partial^{\Delta_a+\Delta_b-1}\delta(x-y)
\nonumber\\ &+& \sum_c
f^{ab}_{(1)}{}_c[\partial^{ \Delta^a + \Delta^b-\Delta^c-1 }\delta(x-y)
W^c(y) +
g^{ab}_c\partial^
{ \Delta^a + \Delta^b-\Delta^c-2 }\delta(x-y) \partial W^c(y) +\ldots ]
\nonumber\\
&+& \sum_{c,d} f_{(2)}^{ab}{}_{cd}\partial^{ \Delta^a +
\Delta^b-\Delta^c-\Delta^d-1 }
\delta(x-y) [  W^c(y)W^d(y) + \ldots ] +
\ldots,
\label{eq.stcl}
\enn
where the right hand side has
been ordered according to degree in $W$,
and $f^{ab}_{(i)}{}_{c..z}$ may be
functions of the central charge $c$.
Alternatively, if we take the space on which the fields are
defined to be the unit complex circle, we can expand the fields in
modes exactly as in the quantum case. Identical mode algebras are
generated if, in the equations (\ref {eq.stqu}), (\ref {eq.stcl})
we use the correspondence
\eq
\zz{-N} \to (-1)^{N-1}2\pi i\delta^{N-1}(z-z')/(N-1)!
\;,
\label{eq.corropepb}
\en
although if we use identical structure constants we do not expect
that both quantum and classical commutator algebras satisfy the Jacobi
identity if they are
non-linear.
We shall assume that the classical algebras have a number of properties
that they would inherit automatically as the classical limit of the
quantum algebras. They contain a classical version of the Virasoro
algebra
\eq
{1\over {2\pi }}\{L(x),L(y)\}=-{c\over 12}\delta'''(x-y)
-2 L(y)\delta'(x-y)+
\partial L(y)\delta(x-y)
\;,
\label{eq.virclass}
\en
and the generating fields obey the classical version of (\ref {eq.pr}).
Further, the only terms in the Poisson bracket algebra which are
independent of $W^a$ are taken to be of the form
$m(m^2-1)\cdots(m^2-(\Delta_a-1)^2)\delta_{m+n,0}\delta^{ab}$.
By analogy, we
refer to such algebras as {\it positive-definite classical} algebras.

We now discuss the
relationship between a classical $W$-algebra and its quantised version.
We shall see that an
extremely important criterion for a quantum algebra to have a
classical limit is that it is well-defined for all values of the
central charge $c$, with the exception perhaps of a few isolated
values or closed intervals. By this we mean that that there is an
operator product algebra for a set of fields $W^a(z)$ of fixed
conformal weights $\Delta_a$, with structure constants $f_{(i)}$ which
are continuous functions of $c$, which is associative for a continuous
range of $c$ values.
We call such algebras {\it deformable}. We shall see that deformability
is however not sufficient for a quantum algebra to have a classical
limit.

As an example let us first consider the classical Virasoro algebra,
(\ref{eq.virclass}).
Quantising this algebra yields
\eq
{}~[ L'_m, L'_n] = \hbar {c'\over 12}m(m^2-1)\delta_{m+n,0} + (m-n)\hbar
L'_{m+n}
\;,
\en
where the prime ${}'$ indicates that we have the normalisation
inherited from the classical Poisson bracket structure. To recover the
standard normalisation we must substitute
\eq
L' = \hbar L \quad, \qquad c'= \hbar c
\;.
\en
Similarly, for a general quantum W-algebra, we can
re-introduce $\hbar$ by the substitutions
\eq
L \to L'/\hbar, c \to c'/\hbar, W^a \to W^a{}'/\hbar^{\alpha_a}
\;,
\label{eq.sub}
\en
where the constants ${\alpha_a}$ are to be determined.
The classical limit is given by the usual correspondence
\eq
\{W^a,W^b\} = \lim_{\hbar \to 0}{ 1\over i\hbar }[W'^a,W'^b].
\en
For this limit to make sense we require that the quantum operator
product algebra remain
associative as $\hbar \to 0$, or equivalently, as $c\to \infty$.
This is why the $W$-algebra must be deformable.
Substituting (\ref{eq.sub}) into (\ref{eq.stqumodes2}) we obtain schematically
\eq
{}~[W_a',W'_b] = c' \hbar^{2\alpha_a -1}\delta
+ f_{(1)} W'_c \hbar^{\alpha_a+\alpha_b -\alpha_c}
+ f_{(2)} \mno W'_e W'_f \mno \hbar^{\alpha_a + \alpha_b - \alpha_f -
\alpha_e} + O(W^3)
\;.
\label{eq.c2}
\en
For this to be the quantisation of a classical W-algebra, we require
that the right-hand side be $O(\hbar)$. The $\hbar$ dependence comes
both explicitly from $W \to W' \hbar^\alpha$, but also implicitly,
from $f_{(i)}(c) \to f_{(i)}(c'/\hbar)$.
If we have
\eq
f^{ab}_{(i)\{c_i\}} = O(\hbar^{\gamma_i(a,b,\{c_i\})})
\hbox{ as $\hbar \to 0$}\;,
\en
and all the fields $\mno (W')^p\mno$ are $O(1)$, then we must impose
that for each term in the singular part of the operator product
algebra expansion of $W^a$ and $W^b$
\eq
\min_{(a,b,c_i)}( \gamma^i  + \alpha_a + \alpha_b - \sum_{j=1}^i
\alpha_{c_i} ) \ge 1
\;.
\label{eq.c4}
\en
If it is not possible to find such constants $\alpha_a$, then we say
that the W-algebra has no classical limit.
The restriction (\ref{eq.c4}) only restricts the couplings to fields
which appear in the commutation relation, or equivalently to
fields in the  singular part of the operator product of
$W^a$ and $W^b$. It is obvious that we have no restriction on the
regular terms since
\eq
W^a(z) W^b(z') = \ldots + \no:W^a(z')W^b(z'): + \ldots \;\;,
\en
where the coupling to $\no:W^aW^b:$ is $O(1)$.

If the classical limit of a W-algebra is positive-definite, we call the
quantum algebra {\it reductive}. To examine what restrictions this implies,
we must consider the central terms in (\ref{eq.c2}).
Fixing the behaviour of the central term
we require
\eq
\alpha_a = 1 \hbox{ for all $a$}
\;.
\label{eq.c1}
\en
For this choice of $\alpha_a$ the requirement (\ref{eq.c4}) becomes,
for each term which appears in the singular part of the operator
product expansion,
\eq
f_{(i)} = O(c^{1-i}) \hbox{ as $c \to \infty$.}
\label{eq.c3}
\en

If it is not possible to impose (\ref{eq.c4}), then the only
possibility of recovering a classical W-algebra is that the normal
ordered products are no longer $O(1)$. Generically for a W-algebra
which comes from the quantisation of a classical algebra we have
\eq
\mno W^a(x) W^b(y) \mno = W^a(x) W^b(y) + O(\hbar)
\;.
\en
It is possible for the first term to vanish if the bosonic fields $W^a,W^b$
can be written as composite fermionic fields,
\eq
W^a  = d(x) f(x) \quad,\qquad W^b = d(x) e(x)
\;,
\en
where classically $d(x)d(x) = 0$, and quantum mechanically
$\mno d(x) d(y) \mno = O(\hbar )$.
A simple example of this possibility may be seen by
considering the first two  W(4,6) algebras of ref  \cite{KWat1}. The
coupling constants of these algebras do not meet the
requirements (\ref{eq.c3}) or even (\ref{eq.c4}).
In fact one of these two algebras can be constructed as the
bosonic `reduction' of a fermionic W-algebra \cite{Bouw3}, the $N=1$
superconformal algebra. This yields a W-algebra with fields of spins 4
and 6, with zero central charge classically. We shall discuss such
reductions further in section 5.

Let us now turn to the question of constructing the advertised
finite Lie algebra from classical and quantum $W$-algebras.
Although the full set of modes of a quasi-primary operator $\cO^i$
only form an indecomposable representation of $su(1,1)$, the subset of modes
\eq
\{ \cO^i_m \;:\; \mod m < \Delta(\cO^i) \}
\label{eq.vpm}
\en
form an irreducible representation of $su(1,1)$.
The set of all such modes for a $W$-algebra forms a closed subalgebra.
We call this the {\it vacuum-preserving algebra} (vpa), since
in the quantum case these are precisely the modes which annihilate both
the right and left $su(1,1)$ invariant vacua.
Although this algebra involves only a finite number of modes of each field,
for a non-linear algebra it will
only close on the modes associated with an infinite
number of such quasi-primary fields, so that it is not a finite Lie algebra.
For the linear Virasoro algebra the vpa is the set $\{L_1,L_0,L_{-1}\}$ which
form the algebra $su(1,1) = sl(2,\Bbb R)$;  for the superconformal
algebra, the vpa is the algebra $osp(1,2)$.
These subalgebras
give useful information about the structure constants of fields in
conformally invariant and superconformally invariant theories
respectively and we would like to define a similar finite Lie algebra
associated to a general W-algebra.

Let us consider first a classical W-algebra. The assumption of
\posdefness\ says that the algebra takes the form
\eqq
\{W^a_m, W^b_n\} &= &
(c/((2\Delta_a -1)! \Delta_a)
 m(m^2-1) \ldots (m^2-(\Delta-1)^2) \delta^{ab}\delta_{m+n} \nonumber\\
&+& f^{ab}_{(1)}{}_c P(\Delta_a,\Delta_b,\Delta_c,m,n) W^c_{m+n} + \ldots
\label{eq.mv}
\enn
If we restrict attention to the vpa, we see immediately that the
central terms are absent. We can now consider a new bracket
$\{ . , . \}_P$ on the modes in the vpa, which consists simply of the
linear term in (\ref{eq.mv}). We can easily check that the Jacobi
identity is satisfied by this new bracket when we restrict to the
vpa.
This is because for $|m|<\Delta^a$ $P(\Delta_a,\Delta_b,0,m,n)=0$ and
consequently
\eqq
\{W^a_m, \sum_p W^b_{n+p} W^c_{-p}\} &=&
\sum_{p} f_{(1)}^{ab}{}_dP(\Delta_a,\Delta_b,\Delta_d,m,n+p)
W^d_{m+n+p}W^c_{-p}
 \cont + f_{(1)}^{ac}{}_dP(\Delta_a,\Delta_c,\Delta_d,m,-p)
 W^d_{m-p}W^b_{n+p} \cont  + O(W^3)
\;,
\enn
so that in the classical case
the contributions to the Jacobi identity from the quadratic and
higher order terms which we have neglected do not contribute, and so
the restricted bracket
\eq
\{W^a_m, W^b_n\}_P = f^{ab}_{(1)}{}_c P(\Delta_a,\Delta_b,\Delta_c,m,n)
W^c_{m+n}
\;.
\label{eq.r1}
\en
is a closed Lie algebra, $g$.
Since we have, by assumption, included the modes $L_{\pm 1},L_0$ in the
vpa, we see that we automatically have an $su(1,1)$ embedding
$su(1,1)\subset g$ given by a classical W-algebra.
In the case of the Zamolodchikov algebra $\WA_2$, we have
the modes $\{Q_{\pm2}, Q_{\pm 1}, Q_0, L_{\pm1}, L_0\}$ forming the
algebra $sl(3)$ with $su(1,1)$ in the maximal regular embedding.

We should like to attempt the extension of this argument to
the quantum case. In the classical case, the contribution
from composite terms to the linear terms in the double commutator vanished for
the vpa because the only possible `contraction' from three fields to one arose
from the central term which decoupled precisely for these modes.
However, in the quantum case there are other contributions to this term which
arise from the need to normal order composite fields. As an illustrative
example we return to the Zamolodchikov algebra $\WA_2$,
this time in its quantised version.
The quantum commutation relations are
\begin{eqnarray}
 \comm {L_m}{Q_n} &=& (2m-n)Q_{m+n}  \label{eq.tqaa}\\
 \comm {Q_m}{Q_n} &=& {c\over3}{{m(m^2 -1)(m^2 -4)}\over{5!}}\delta_{m+n}
		 \nonumber\\
&& + {{(m-n)}\over{30}}(2m^2 - mn + 2n^2 - 8)L_{m+n}
   + \beta (m-n)\Lambda_{m+n} \,,\label{eq.tqa}
\end{eqnarray}
where
\eq
\begin{array}{rclrcl}
\beta &= &{{16}\over{22 + 5c}}\;, &
\Lambda(z) &=& \no:T(z)T(z): %- {3\over 10}\partial^2T(z)
\;.
\\
\end{array} \nonumber
\en
The only non-trivial double commutators are $[L_p[Q_m,Q_n]],[Q_p[Q_m,Q_n]]$
and the
only composite field appearing in the intermediate channel is $\Lambda$, so
we need only consider the linear terms in $[L_p,\Lambda_{m+n}]$,
$[Q_p,\Lambda_{m+n}]$. These are
\eq
{}~[L_p,\beta \Lambda_{m+n}] = \frac{16}{5} \frac{p(p^2-1)}{3!} L_{m+n+p} +
\cdots
\en
%and that
\eq
[Q_p,\beta \Lambda_{m+n}] =
 \beta \frac{^{4(5p^3-5p^2(m+n)+3p(m+n)^2-(m+n)^3-17p+9(m+n))}}{35}
Q_{p+m+n}+\cdots
\en
The first of these commutators vanishes when we take $p=-1,0,1$.
In the second commutator there is a contribution from the $\Lambda$ term in
$[Q,Q]$ which
does not vanish even when we restrict to the vpa, and a more careful
consideration of this term shows that it arises from the need to normal
order composite fields. Instead we can ensure that
this term does not violate the consistency of the `linearised' Jacobi identity
by taking the limit $c\to\infty$. In this case $\b \to0$ and
the vpa linearises, again to give $sl(3,\Bbb R)$.
Note that we need to combine the limit $c\to \infty$ {\it and} the
restriction to the vpa to ensure that both commutators vanish.

We are now in a position to prove this feature, namely that the vpa
linearises to give a finite Lie algebra as $c\to\infty$, assuming that the
W-algebra is reductive.
Let us denote a generic composite field composed of $i$ basic fields
as $\no:(W)^i:$.
The contribution to the Jacobi identity from the coupling through such
terms in
$[W^a,[W^b,W^c]]$ is
\eqq
{}~[W^a_m,[W^b_n,W^c_p]]
	&&= \sum_i [W^a_m, f^{bc}_{(i)}
P(\Delta_b,\Delta_c,\Delta_{(i)},n,p)
\no:(W)^i:_{n+p} ]
\;.
\enn
Let us also write the linear part of the contribution from the
commutator of $W^a$ with $W^{(i)}$,
\eq
{}~[W^a_m, \no:(W)^i:_n] = g^{ae}_{(i)}
P(\Delta_a,\Delta_{(i)},\Delta_e,m,n)
W^e_{m+n} \;.
\label{eq.th2}
\en
At this point we must split our argument into two cases, depending on
whether
$\no:(W)^i:$ appears in the singular or regular part of the operator product
expansion of $W^a$ with $W^e$.
If $\no:(W)^i:$ appears in the singular part of the operator product
expansion then we can calculate the order of
$g_{(i)}$ by taking the three point
function
\eq
C_{abi} = \langle W^a W^e \no:(W)^i: \rangle
\;.
\en
This can be written in two ways. We have that
\eq
C_{abi} = f^{ae}_{(i)} \langle \no:(W)^i:\no:(W)^i: \rangle = O(c)
% \sim c
\;,
\en
using (\ref{eq.c3}) and evaluating the leading contribution in $c$ to
$\cev{0} (W)^i_{\Delta_{(i)}}(W)^i_{-\Delta_{(i)}} \vac$. We also obtain
\eq
C_{abi} = g_{(i)} \langle W^a W^e \rangle \sim \, g^{ae}_{(i)} \, c\;,
\en
using (\ref{eq.stqumodes2})
and (\ref{eq.th2}),
and suppressing non-zero constants. Thus we see that
$g^{ae}_{(i)} = O(1)$
and so the contribution to the Jacobi identity of three basic fields
$[W^a,[W^b,W^c]]$ to the field $W^e$
from the term in $[W^b,W^c]$ of form $\no:(W)^i:$ is
$O(g^{ae}_{(i)}f^{bc}_{(i)})=O(c^{1-i})$, if the
field $\no:(W)^i:$ appears in the singular part of the operator product
expansion of $W^a$ with $W^e$. However, if the field $\no:(W)^i:$ does
not, then we cannot apply (\ref{eq.c3}) to deduce the order of the
coupling $g^{ae}_{(i)}$. In this case we have that $\no:(W)^i:$ has
conformal weight $\Delta_{(i)}\geq \Delta_a+\Delta_e$.
However, the polynomials $P(\Delta_a,\Delta_{(i)},\Delta_e,m,n)$
vanish identically if $|m|<\Delta^a,
\Delta_{(i)} \ge \Delta^a + \Delta^e$ (see ref.  \cite{Bowc2}), and we
can use this
fact to bypass our ignorance of the coupling $f^{ae}_{(i)}$.

This shows that if we consider Jacobi's identity for the vpa
modes of the generating fields
in the limit  $c \to \infty$, then all contributions to linear terms from
composite fields in the intermediate channel drop out.
Since the commutator algebra is a Lie
algebra for all $c$ values by the assumption of deformability, the
only obstruction to the vpa algebra restricted to the generating fields
satisfying the Jacobi identity was from such contributions.
Thus, this algebra in the
$c \to \infty $ limit of a \reductive{\compact\  W-algebra }{semisimple
W-algebra with a classical limit } is a finite Lie algebra.

We have now shown how to recover finite Lie algebras from
positive-definite classical W-algebras and reductive
quantum W-algebras.
We shall call this the {\it linearised} vpa algebra.
\reductive{By the Levi-Malcev theorem, the}{The}
most general form for a finite Lie algebra would be the
semidirect product of a  semi-simple  Lie algebra with
its radical, which is its maximal solvable ideal.
However, we can use the positive-definitness of the classical algebra to
show that the maximal solvable ideal of the finite Lie algebra we have
constructed is in fact its centre, or, in other words, that the
linearised vpa is
the direct sum of a semisimple Lie algebra with an abelian Lie
algebra (For results on the structure of Lie algebras used here, see
e.g.  \cite{Vara1}).

To do this, let us consider the maximal solvable ideal $a$ of the
linearised vpa of a reductive W-algebra.
Let us suppose that a particular mode $W^a_m \in a$.
The modes $L_{\pm 1},L_0$ are always in the linearised vpa, and we know
the commutation relations of $W^a_m$ with $L_m$ to be of the form
\eq
{}~[L_m,W^a_n] = ((\Delta_a-1)m - n) W^a_{m+n} \;.
\label{eq.lw}
\en
Since $a$ is an ideal, eqn. (\ref{eq.lw}) implies that $W^a_m \in a
\Rightarrow W^a_n \in a\;\;$ for all $|n|< \Delta_a $.
With the standard normalisation for a  \posdef\   quantum W-algebra we have
\eq
W^a(z) W^a(z') = (c/\Delta)\zz{-2\Delta_a} + 2 L(z')\zz{-2\Delta_a + 2}
+ \ldots \;.
\en
Using the Virasoro Ward identities, (see e.g. appendix B of ref.
 \cite{BPZ}), we can deduce the coefficients of all the terms
$\partial^i L(z')$ in this operator
product and we can deduce that
\eq
{}~[W^a_{\Delta -1}, W^a_{1-\Delta} ] =
	12(\Delta-1)/(\Delta (2\Delta -1)) L_0 + \ldots
\en
%where $p$ is some polynomial in $\Delta_a$.
For $\Delta_a > 1$ this is non-zero,
and so, for $\Delta_a>1$ we have $L_0\in a$. If $L_0 \in a$, then we
immediately  we get that $L_0,L_{\pm 1}\in a$.
which is a contradiction since a solvable ideal cannot contain
a semi-simple algebra.

So, if $W^a_m \in a$, where $a$ is a solvable ideal, then $\Delta_a \le
1$. If $\Delta_a < 1$ then it contributes no modes to the vpa; if
$\Delta_a=1$ then $m=0$, and
we can thus denote the elements of $a$ as $U^i_0$, the zero
modes of a set of weight one fields $U^i(z)$. These zero modes $U^i_0$
form a solvable Lie algebra.
However, it has been known for a long time (see e.g.  \cite{Godd1})
that the requirement that the inner product on the primary fields of
weight one is positive definite forces them to have a Kac-Moody
algebra based on a compact semi-simple Lie algebra plus some $u(1)^n$
current algebra. Thus the zero modes of weight one fields in cft form a
finite dimensional Lie algebra which is the direct product of a semisimple
Lie algebra with an abelian algebra, and we see that the ideal $a$ is
abelian.

The only possibility left open to us now is that the linearised vpa has the
structure of a semi-simple Lie algebra semidirect product with an
abelian algebra.
Suppose that $U_0 \in a$. Then $[U_0,W^a_m] \in a$ for all $W^a_m$
in the vpa.
If $[U_0,W^a_m] = X_m$, then the operator product expansion of $U(z)$
with $W^a(z)$ must be of the form
\eq
U(z) W^a(z') = \ldots + X(z') \zz{-1} + O(1) \;,
\label{eq.uwx}
\en
since $U_0 = \int dz/(2\pi i) U(z)$. However, from (\ref{eq.uwx}) we
see that the field $X$ must have conformal weight equal to that of
$W^a$. Since $X_m\in a$, we see that $X$ must have weight one, and so
$W^a$ must have weight one. We already know that $a$ commutes with the
zero modes of the spin one fields, so in fact $a$ is the centre of the
vpa.
This completes the proof that the linearised vpa of the classical limit of
a reductive W-algebra is the
direct sum of a semisimple Lie algebra with an abelian algebra.

The above discussion has been for a purely bosonic W-algebra. If we
wish to include fermions then we must also consider Lie superalgebras,
since the vpa of fermionic fields will contain anti-commutators of the
modes of fermionic fields.
We shall use the notation of  \cite{Kac3} for Lie superalgebras, with
the algebra decomposition
\eq
g = g_{\bar 0} \oplus g_{\bar 1}
\en
where the bosonic generators are in $g_{\bar 0}$ and the fermionic in
$g_{\bar 1}$.
We define the grade of a generator $X$ to be $g(X) = j$ if $X\in
g_{\bar\jmath }$.
The (super)Lie bracket then takes the form
\eq
{}~[ X, Y] = XY - (-1)^{g(X)g(Y)} YX
\en
The bosonic fields have modes in $g_{\bar 0}$ and the fermionic fields
have modes in
$g_{\bar 1}$. It will also be the case that the bosonic fields will
have integral conformal weight and the
fermionic fields half-integral conformal weight, for unitarity.
A fermionic field of weight $\Delta$ will have mode decomposition
\eq
\psi(z) = \sum_{n\in Z + 1/2} \psi_n z^{-n-\Delta}
\en
As for bosonic fields, the vpa contains the modes of the fermionic fields
\eq
\psi_m \;: |m| < \Delta
\en
We see that for a free fermionic field of conformal weight $1/2$ ,
there are no modes in the vpa. Thus analysis of the vpa will yield no
information on the free fermion content of a theory. However, this is
not obstacle since free fermions have already been shown to factorise
from the Hilbert space by Goddard and Schwimmer \cite{GSch2}.

Since the  fermionic fields will have half
integer modes, the $su(1,1)$ decomposition must be compatible in the
sense that the decomposition takes the form
\eq
g_{\bar 0} = \oplus_{j \in Z} D_j \;,\;\; g_{\bar 1} = \oplus_{j\in Z+
1/2} D_j
\label{eq.grad}
\en
where $D_j$ is the representation of dimension $2j+1$.
We can also prove that the superalgebra consists of the direct sum of
simple (super)-algebras and an abelian Lie algebra in an analogous
manner to that above for purely bosonic W-algebra vpa's.

This means that the field content of any {\it  \posdef\  } W-algebra
which is defined for all $c$ values must comprise a set of free
fermion fields of weight one half (which do
not contribute to the vpa; such fields have already been shown to
factorise  \cite{GSch2}), a set of
bosonic free fields of weight one ($u(1)^n$ current algebra) and a set
of fields whose weights are given by an $sl(2)$ embedding in
a semisimple Lie superalgebra which is compatible with the grading of
the superalgebra as in (\ref{eq.grad}).

For a purely bosonic W-algebra, the field content will comprise a set
of weight one fields and a set of bosonic fields whose conformal spins
are given by an integral $su(2)$ embedding in a semi-simple Lie
algebra.

We shall now go on to show that this is indeed the case for the
generalised Drinfeld Sokolov constructions mentioned earlier, and then
to consider various cases of particular interest. The rest of this
paper will be concerned only with the case of bosonic algebras for
simplicity.

\def\mychoice#1#2{#2}

\section{Hamiltonian systems and co-adjoint orbits}
\label{sec.hs}
\label{sec.u10}
\def\rhov{{\rho^\vee}}
The analysis of the previous section showed that to each
 \reductive{\compact\ }{semi-simple} W-algebra one could associate
a finite Lie algebra with some $su(1,1)$ embedding specified.
This is reminiscent
of the data that is required for a generalisation of the
Drinfeld-Sokolov
construction of Hamiltonian structures that have been studied recently
\cite{Int,DUB2,PRIN}.
In the next section we
shall show that this data is recovered as the finite Lie algebra we
constructed. This provides us with an existence proof for the
classical W-algebras associated with each Lie algebra and $su(1,1)$
embedding. In this section we give a brief review of this construction.

The classical Hamiltonian systems of Drinfeld and Sokolov \cite{DSok}
are based on
a Poisson bracket structure on $g^*$, the dual to the Lie algebra $g$,
and the extension of this to $\hat g$, the centrally-extended
Kac-Moody algebra related
to $g$. An element of $\hat g$ consists of a pair,
\eq (j(z),c)\,,
\en
where $c$ is a number and $j$ is a field on $S^1$ valued in $g$.
The coordinate on $S^1$, we denote by $z$, with $0\le z<2\pi$.
%(For a brief, readable account of the coadjoint method and its
%connection with R-matrix theory, see \cite{Seme1}.)
With this definition, the Lie bracket of two  elements of
$\hat g$ is given by
\eq
[(j^1(z),c^1),(j^2(z),c^2)]
 = ([j^1(z),j^2(z)],k\int\Tr\{\partial j^1(z)j^2(z)\}\,dz )
\;,
\en
where the second term corresponds to the cocycle of $\hat g$.
An element of $\hat g^*$ is given by a pair $(q,\lambda)$, where $q$
is a $g$-valued field on $S^1$ and $\lambda$ is a number.
%There is a group invariant form on $\hat g$ given by
The action of this element on $(j,c)$ in $\hat g$ is given by
\eq
\ip{(q,\lambda)}{(j,c)} = \int\Tr \{qj\} \,dz + c\lambda \;.
\en
With this, we may identify $\hat g$ and $\hat g^*$, and we obtain a
canonical action of $\hat g$ on $\hat g^*$, the coadjoint action $ad^*$.
If $(q,\lambda )\in\hat g^*,\,(j^i,c^i)\in\hat g$, then we have
\begin{eqnarray*}
ad^*_{(j^1,c^1)} \cdd (q,\lambda ) [(j^2,c^2)]&=& (q,\lambda
)[ad_{(j^1,c^1)}\cdd (j^2,c^2)] \\
	&=& \ip{(q,\lambda )}{([j^1,j^2], k\int \Tr\{(j^1)'j^2\})} \\
	&=&\int \tr \{[q,j^1]j^2 + k\lambda \partial j^1 j^2\}\,.\\
\noalign{\ni Thus we obtain}
ad^*_{(j^1,c^1)}\cdd (q,\lambda ) &=& ([q+k\lambda\partial,j^1] ,0)\,.\\
\end{eqnarray*}
This is simply an infinitesimal gauge transformation of $q$.
This phase space also has a canonical action of $\hat G$, the
coadjoint action $Ad^*$, given by
\eqq
Ad^*_U\cdd q (h) &=& q (Ad_U\cdd h)\,, \\
Ad^*_U \cdd (q,\lambda) &=& (U^{-1}qU + k\lambda U^{-1}\partial U,\lambda)\,.
\enn
There is a canonical Poisson bracket structure on $\hat g^*$, the
Berezin-Kirilov-Kostant-Lie-Poisson bracket. If $U,\,V$ are two
functionals on $\hat g^*$, then their Poisson bracket is also a functional on
$\hat g^*$. When evaluated on $q\in\hat g^*$ it is explicitly given by
\eqq
\{U,V\}_{q} &=& \ip{q}{[d_qU,d_qV]_{KM}} \;,%\nonumber\\
%	&=& \int\Tr(j[d_jU,d_jV]) +c\Tr(d_jU'\,d_jV)\,dz\;,
\enn
where $d_jU,d_jV$ are any elements of $\hat g$, such that for all
$\delta j\in \hat g^*$,
\eq
U(j + \delta j) = U(j) + \ip{d_jU}{\delta j} + O(\delta j^2)\,.
\en
We shall usually suppress the $j$ suffix if it is clear from context.

We may accordingly evaluate the Poisson brackets of the
components of $\hat g^*$.
If $\{T^i\}$ form a basis of the generators of $\hat g$ with
$\Tr\{T^iT^j\}=g^{ij}$, then we can define the functionals $\hat
T^i(x),\hat e$ on $(q,\lambda)\in\hat g^*$ by
\eq
\begin{array}{rcl}
\hat T^i(x)[(q,\lambda)] &= & \Tr\{T^i q(x)\} \,,\\
\hat e[(q,\lambda)]	&=& \lambda	\,.
\end{array}
\en
We have
\eq
\begin{array}{rcl}
d\hat T^i(x) 	&=& (T^i \delta(x-y),0) \;,
\\
d\hat e		&=& (0,1) \;.
\end{array}
\en
Thus we can evaluate the Poisson brackets of these functionals
\eqq
\{\hat T^i(x),\hat T^j\}_{(q,\lambda)} &=&
	\int\,dz\,\Tr\{ q[T^i\delta(x-z),T^j \delta (y-z)]
	+k\lambda\partial_z(T^i\delta(x-z))T^j \delta(y-z) \} \;.\nonumber\\
\noalign{\ni Thus we see that}
\{\hat T^i(x),\hat T^j\}_{(q,\lambda)} 	&=&
	f^{ij}{}_k\hat T^k(y)\delta(x-y) - k\hat eg^{ij}\delta'(x-y)
\;.
\enn
This is the Kac-Moody algebra $\hat g$. In particular we shall often
denote the zero grade subalgebra functional $\hat J^i$ by $j^i$ and
$\hat e$ by 1, with the Poisson brackets
\eq
\{j^i(x),j^j(y)\} = f^{ij}{}_k j^k(y)\delta(x-y)
	- \half k\delta^{ij}\delta'(x-y)\;.
\label{eq.hpbs}
\en

The method of hamiltonian reduction involves constraining currents
associated with nilpotent elements of the algebra. In the traditional
reduction associated with Toda theory or the standard KdV hierarchy
one gauged the maximal nilpotent algebra associated with, say, all
the positive roots of $g$. It was then realised that one could generalise
this construction by gauging some smaller set of currents, and
moreover, that this set could be succinctly
labelled by some $su(1,1)$ embedding.
Since we are interested in bosonic positive-definite $W$-algebras,
we may assume that the $su(1,1)$ embedding is integral.
Non-integral
embeddings result in bosonic fields of half-integral weight and the
resulting $W$-algebras are not positive-definite.

Let us consider some modified Cartan-Weyl basis for $g$,
\eq
\mychoice{g=g^{\alpha_-}\oplus h\oplus g^{\alpha_+}}{g=g^-\oplus
h\oplus g^{+}}
\;.
\en
Here
\eq
g^\pm = \oplus \Bbb C E^{\pm\alpha} \;, \; h = \oplus \Bbb C H^i
\;,
\en
with the commutation relations
\eq
\comm{E^\alpha}{E^{-\alpha}} = (2/\alpha^2) \alpha^i H^i \;, \;
\comm{H^i}{E^\beta} = \beta^i E^\beta
\;.
\en
One can always conjugate any $su(1,1)$ subalgebra of $g$ so that
\mychoice{$I_+\in g^{\alpha_+},I_0\in h, I_- \in g^{\alpha_-}$
where}{$I_+\in g^{+},I_0\in h, I_- \in g^{-}$ where}
$I_+,I_-,I_0$ are
the usual raising, lowering and diagonal basis of $su(1,1)$.
We may write
$I_0=\rhov\cdot H$.
If we use the standard normalisation for the $su(1,1)$ algebra,
\eq
{}~[I_0,I_\pm]= \pm I_\pm\;,\; [I_+,I_-] = \sqrt 2I_0 \;,
\en
then
we may define the characteristic of the
$su(1,1)$ embedding to be $(\rhov\cdot e_1,...,\rhov \cdot e_i)$,
where $e_j$ are the simple roots of $g$. It is a fact that the
entries of the characteristic are $0,1/2,1$. For integral embeddings
they must either be $0$ or $1$. The standard reduction is associated with
the principal embedding whose characteristic contains all ones.

We may grade $g$ with respect to the $\rhov\cdd H$ eigenvalue as
\eq
g = \oplus_m g_m \,.
\en
The elements of $g$ which are highest weight states for this $su(1,1)$ action
form a commuting subalgebra of $g$. % ( see e.g. \cite{OTur1}).
We denote these highest weights by $E^{(e_i)}$, and the
corresponding lowest weights by $E^{(-e_i)}$. The highest weights
are annihilated by $I_+$ and the lowest weights by $I_-$.
We denote the subalgebra $\oplus_{n\ge 0}g_n$ by $p^+$ and the subalgebra
\mychoice{$\oplus_{n>0}g_n$ by $n^+\subset g^{\alpha_+}$. Similarly
for $p^-,n^-$.}{$\oplus_{n>0}g_n$ by $n^+\subset g^{+}$. Similarly for
$p^-,n^-$.}
%If the generators of the Cartan subalgebra are denoted by the vector
For the standard reduction associated with the principal reduction,
%$n^\pm=g^{\alpha_\pm}$.
$n^\pm=g^{\pm}$.

Since $\hat G$ acts on $\hat g^*$,
we may perform a classical Hamiltonian reduction \cite{Arno1}
with respect to the subgroup
$\hat N^-$, where $N^-$ is the
subgroup of $G$ which has the nilpotent subalgebra $n^-$
as its Lie algebra.
In this procedure one chooses an image of the momentum map $\pi$ and
the phase space consists of equivalence classes under the residual
symmetry of the inverse image of $\pi$. Here $\pi$ is essentially the
projection map $g \mapsto n_+$.
We can choose the image of $\pi$ in such a way that the inverse image
consists of elements of $\hat g^*$ of the form
\eq
(b(z) + I_+,1)\,,
\en
where $b(z) \in p^-$.  We call this space $M$.

The action of $\hat N^-$ is now an equivalence relation on $M$. From
the form of $M$, we may choose coordinates on this space to be gauge
invariant differential
polynomials of the entries in the matrices
$b(z)$. In particular, there is a unique gauge transformation
$Ad^*_N = \exp(ad^*_n)$ with $n \in \hat n^-$
which gives
\eq
N^{-1}(b + I_+)N + kN^{-1}\partial N = I_+ + \sum_n W_n E^{(n)}\,,
\label{eq.partw}
\en
where $E^{(n)}$ span the kernel of $I_-$.
To show this gauge transformation is unique, take components in $g_m$.
As a result, the entries of $n$ are uniquely determined %differential
polynomials in the entries of $b(z)$.

The Poisson bracket structure on gauge invariant functionals
$\phi,\psi$ on classes of $q\in\tilde M$ is given by
\eq
\{\phi,\psi\}_q = \ip{q}{[\nabla_q\phi,\nabla_q\psi]_{KM}}
\;,
\label{eq.wal}
\en
where $j=\nabla_q \phi$ is any element of $\hat g$ such that
\eq
\phi(q+\delta q) = \phi(q) + \ip{\delta q}{\nabla_q \phi} + O(\delta
q^2)
\;,
\en
for all $q \in M$. Thus $\nabla_q \phi$ is determined up to an element
of $\hat n^-$; one such choice is $\nabla_q\phi = d_q\phi$.
It is easy to show that this bracket is well defined
and satisfies the Jacobi identity \cite{DSok}.

Further, we can imbed this structure in the Lie-Poisson bracket
structure by considering the map $\mu$ from $\hat G_0$, corresponding to
zero-graded Lie algebra $g_0$,
to gauge invariant functionals on $M$ given by
\eq
\mu: (g_0+I_+,1) \mapsto \{ W_i(g_0) \} \,,
\en
where
\eq
N^{-1}(g_0+I_+)N + kN^{-1}\partial N = I_+ + \sum_n W_n E^{(n)} \,.
\en
Since this gauge
transformation is unique, the %differential
polynomials $W_n(g_0)$ are a
choice of coordinates on the manifold $\tilde M =M/\hat N^-$.
It can also be shown \cite{DSok} that the Poisson bracket structure
on $\hat g$ and $\tilde M$ are compatible in the sense that
\eq
\{ \mu^*\phi,\mu^*\psi\} = \mu^* \{\phi,\psi\}\,,
\en
where $\mu^*\phi$ is a functional on $\hat g^*$. $\mu^*$ is called the
generalised Miura transformation. For the standard reduction corresponding to
the principal embedding, $g_0=h$, and so the Miura transformation
provides a free field representation of the
Poisson bracket algebra (\ref{eq.wal}). For more general reductions we
obtain a construction in terms of the
currents of the zero-graded (non-abelian) algebra.
Since the polynomials $W_n(g_0)$ are a choice of coordinates on $\tilde M$,
this
algebra is closed, although not necessarily on linear combinations of the
original coordinates.

\section{Linearised Poisson brackets for classical $W$-algebras}

We are now in a position to calculate the Poisson brackets of the
W-algebra given by the
Drinfeld-Sokolov reduction associated with some Lie algebra $g$ and
a particular integral $su(1,1)$ embedding. The purpose of this section is
to show that the finite subalgebra and $su(1,1)$ embedding of section 2
associated with this W-algebra coincide with those chosen to specify the
reduction. This will demonstrate the existence of the classical W-algebra
associated to each such pair.
We can check this in this case by using the expressions we
deduced in the previous section
for the W-algebra Poisson brackets. The finite subalgebra is simply
the `linearised' vpa for the generating fields,
so we will only need to calculate the Poisson
brackets to linear order in the fields. This is the feature which makes
the calculation tractable.

We need to calculate the Poisson brackets of the functionals $\cH$,
\eq
\cH = \int \dz f^a(z) W^a(z)
\;,
\en
where %and
$W^a$ is a W-algebra field and now $z$ is a complex coordinate.
\blank{
 with
\eq
\int \dz = \oint_0 \frac{dz}{2\pi i}
\en
}
If we denote the W-algebra fields of weight $\Delta_a = a+1$ by $W^a$,
then the modes $W^a_m$ are given by $\cH$ for $f(z) = z^{m+a}$.
We shall not differentiate between fields of equal conformal weight,
to avoid proliferation of indices. From (\ref{eq.partw}) we know that
the gauge invariant fields correspond to the lowest weights of the
$sl(2)$ embedding. If we wish to identify a particular field of a
given weight, then we shall use the notation $W[X]$, where $X$ is a
generator of $g$ which is a lowest weight of the $sl(2)$.

The results of the last section tell us that the Poisson bracket
algebra is given by
\begin{eqnarray}
\{W^a_m, W^b_n\} &=&
	\int \dz \Tr( j(z) [dW^a_m, dW^b_n] ) + k \int \dz \Tr(dW^a_m
(dW^b_n)')
\;.
\label{eq.d2}
\end{eqnarray}
If we are interested in the term in this Poisson bracket which is
linear in the fields $W$, then we clearly only need to calculate
$dW^a_m$ to linear order in the fields $W$.

Consider the arbitrary element of the space $M$ to be of the form
\eq
j = b(z) + I_+ \;.
\en
We may decompose $g$ with respect to the $sl(2)$ subalgebra to find
the highest weights of this $sl(2)$ which we shall denote by $E^a$.
Then bases for $g, g^-$ and $n^-$ are given by
\eqq
g &=& \oplus_a \oplus_{i=0}^{2a+1} \Bbb C E^{i,a} \;,\\
g^- &=& \oplus_a \oplus_{i=0}^{a+1} \Bbb C E^{i,a} \;,\\
n^- &=& \oplus_a \oplus_{i=0}^a \Bbb C E^{i,a} \;.
\enn
where
\eq
E^{i,a} = ad^i(I_+) \circ E^a \;.
\en
Then we may write
\eq
j  = I_+ + j_0 +
I_+ + \sum_a \sum_{i=0}^a j^{i,a}(z) E^{i,a} \;.
\en
We shall now consider a gauge transformation of the form
\eq
j^l =
\exp(ad(l)) \circ ( j + \k\partial)  \;,
\label{eq.d1}
\en
where $l$ is some current in
%$g^{\alpha-}$.
$n^-$.
If $l$ is the transformation which puts $j$ into the highest weight gauge,
then  $l$ is defined implicitly as a polynomial function in the
components of $j_0$ and their derivatives, by
\eq
j^l = j^W =I_+ + \sum_a W^a E^a \;.
\label{eq.dd}
\en
An important step in the argument is to decompose $l$ into components
which are homogeneous in the components of $j_0$, and further into
components of $E^{i,a}$:
\eqq
l &=& \sum_j l^{(j)} \\
	&=& \sum_j \sum_a \sum_{i=0}^{a-1}  l^{j|ia} E^{i,a} \;,
\label{eq.ddd}
\enn
where $l^{p|rb}$ is homogeneous in $j_0$ of degree $p$.
If we decompose the gauge invariant functionals $W^a$ into
homogeneous pieces $W^{p|a}$ of degree $p$, then upon
substituting (\ref{eq.ddd}) into eqn. (\ref{eq.dd}), we obtain
\eqq
W^{1|a} &=& j^{0,a} - \k(l^{1|0a})' \\
	% - \k(l^{2|0a})' + \ldots\;,\\
W^{2|a} &=&
\Tr \left( E^{2a,a}[l^{(1)},(  j_0 + \half 1 [l^{(1)} ,I_+] )]  \right)
- k (l^{2|0a})' \;,\\
\noalign{\noindent where}
l^{1|i,a} &=& \sum_{p=i+1}^{a} (-\k\partial)^{p-i-1} j^{p,a} \\
l^{2|i-1,a} &=&
\sum_{p=i}^{a-1} (\k\partial)^{p-i}(-1)^p\left[
\Tr\left(E^{2a-p,a}[ l^{(1)}, (j_0 + \half 1 [l^{(1)}, I_+])]\right)
\right] \;.
\enn
We have chosen a normalisation for the $sl(2)$ highest weight vectors
\eq
\Tr( E^{2a,a} E^{0,a} ) = 1\;,
\label{eq.lnorm}
\en
For simplicity we have used notation which assumes that there is a unique
field of each weight, but the generalisation is straightforward. The
normalisation (\ref{eq.lnorm}) means that the generators $I_\pm,
[I_+,I_-]$ obey $su(1,1)$ commutation relations with nonstandard
normalisation factors.

We can now deduce  $d\cH$ for $\cH = \int W^a(z)
f(z) \dz$. This will in general be  a function of the entries $j^{ia}$,
and since we are interested in the Poisson brackets of gauge-invariant
quantities, we may simply substitute $j$ by $j^W$, after we have
evaluated $dW$. This makes the evaluation of $dW$ very easy since
$l^{i|ja}=0, j^{ia}= \delta^i_0 W^a$. We can thus identify the terms
which will contribute to $d\cH$ where $\cH = \int f W^a$.
Using the notation
\eq
\left\{\matrix{a&b&c\cr j&k&l}\right\} =
\Tr\{ E^{j,a} [E^{k,b}, E^{l,c}] \}
\;,
\en
we have
\eqq
 d \cH &=&
\int  \sum_{p=0}^a (\k\partial)^p f dj^{i,a} \nonumber\\
&+&
\int  \sum_{p=0}^{a-1} \sum_{b,c} \sum_{q=0}^{b-1} \sum_{r=q+1}^b
(\k\partial)^{r-q-1} \left[ W^c (-\k\partial)^p f \right]
\cul a,b,c,2a-p,q,0. dj^{r,b}
\;.
\enn
Using $\Tr( j dj^{i,a}) = j^{i,a}$, we can easily see that
$dj^{i,a} = (-i)^i E^{2a-i,a}$. We are really only interested in
the case $f = z^{a+m}$ where %$|m| \le a$. In this case
$\cH = W^a_m$, and so finally we obtain
\eqq
dW^a_m &=& \sum_{i=0}^a (a+m)_i (-\k)^i z^{m+a-i} E^{2a-i,a} \nonumber\\
	&+& \sum_{b,c} \sum_{i=0}^{a-1}
	\sum_{q=0}^{c-1}\sum_{j=1}^{c-q}
	(-1)^{q+j} \partial^{j-1}(W^b \partial^{i} z^{m+a})
\left\{ \matrix{b&a&c\cr 0&2a-i&m} \right\} E^{2c-j-q,c}\k^{i+j-1}
\nonumber\\
	&+& O(W^2) \;,
\label{eq.dch}
\enn
where we denote
\eq
(b)(b-1) \ldots (b-c+1) \hbox{ by } (b)_c \;.
\en

We are now in a position to evaluate the Poisson brackets of the
modes $W^a_m$ to linear order in the fields.
We shall decompose $dW$ into its homogeneous pieces, as
\eq
dW^a_m = dW^{a(0)}_m + dW^{a(1)}_m + O(W^2)\;.
\en
Using (\ref{eq.d2}) we see that the terms which contribute to the linear
piece of the Poisson bracket are
\eqq
\{ W^a_m, W^b_n \} &=&
	  \k\int \dz \Tr( dW^{a(0)}_m (dW^{b(0)}_n)')
	+ \int \dz \Tr( j^W(z) [dW^{a(0)}_m, dW^{b(0)}_n] ) \\
&	+& \k \int \dz \Tr(dW^{a(0)}_m (dW^{b(1)}_n)')
	+ \k \int \dz \Tr(dW^{a(1)}_m (dW^{b(0)}_n)')
	+ O(W^2)\;.  \nonumber
\label{eq.d3}
\enn
O'Raifeartaigh et al. in \cite{DUB2,Int} have shown that $L = \alpha
W[I_-] + \sum_a \beta^a W[U^a]^2 $ is a Virasoro algebra for the system
we have described, where $W[I_-]$ is the field corresponding to the
representation of the embedded $sl(2)$ itself, and $W[U^a]$ are fields
corresponding to the singlets in the decomposition of $g$ with respect
to the embedded $sl(2)$. The fields $W[E^{0,a}]$ transform as primary
fields of weight $a+1$ w.r.t. this Virasoro algebra. The central term
in the Virasoro algebra is generated purely by the field $W[I_-]$,
since we have already shown that there are no central terms in the
Poisson brackets of composite fields. In the case $W = W[I_-]$,
\blank{
We shall first consider the special case of $W=L$, to show that
there is a Virasoro algebra in the algebra of functionals $W$.
In the highest weight gauge (\ref{eq.dd}) the field $W=L$ corresponds
to the highest weight representation given by the embedded $su(1,1)$
itself.
In this case eqn. (\ref{eq.dch}) becomes exact,
\eq
dL_m = z^{m+1}[I_+,[I_+,I_-]] - k(m+1)z^m[I_+,I_-]\;.
\en
We can now evaluate the Poisson bracket
\eq
\{L_m,L_n\} = \k\lambda(m-n)L_{m+n} + \k^3m(m^2-1)\delta_{m+n}
\;,\label{eq.svir}\en
which corresponds to a rescaled Virasoro algebra. In (\ref{eq.svir})
$\lambda$ is defined by
\eq
[I_+,[I_+,I_-]] = \lambda I_- \;.
\label{eq.ldef}
\en
Thus, if we re-scale $L$ to return (\ref{eq.svir}) to the standard
normalisation, then we recover $c= 12 k/\lambda^2 = 12 k (\rhov)^2$
where $\rhov$ defines the $su(1,1)$ embedding.
}
eqn. (\ref{eq.dch}) becomes exact,
\eq
dW[I_-]_m = z^{m+1}[I_+,[I_+,I_-]] - k(m+1)z^m[I_+,I_-]\;.
\en
We can now evaluate the Poisson bracket
\eq
\{W[I_-]_m,W[I_-]_n\} = \k\lambda(m-n)W[I_-]_{m+n} + \k^3m(m^2-1)\delta_{m+n}
\;,\label{eq.svir}\en
which corresponds to a rescaled Virasoro algebra. In (\ref{eq.svir})
$\lambda$ is defined by
\eq
[I_+,[I_+,I_-]] = \lambda I_+ \;.
\label{eq.ldef}
\en
Thus, if we re-scale $W[I_-]$ to return (\ref{eq.svir}) to the standard
normalisation, we recover $c= 12 k/\lambda^2 = 6 k (\rhov)^2$
where $\rhov$ defines the $su(1,1)$ embedding.

Thus the Poisson brackets (\ref{eq.d3}) represent a W-algebra.
To establish that this is a positive-definite W-algebra,
note that the first term in (\ref{eq.d3}) corresponds to the
central term and is easy to compute;
\eqq
&& \int \dz \sum_{i=0}^a \sum_{j=0}^b (a+m)_i (b+n)_(j+1) (-\k)^{i+j+1}
	z^{m+n+a+b-i-j-1} \Tr(E^{2a-i,a} E^{2b-j,b})\;. \nonumber\\
\noalign{\noindent The trace in this term gives
$\delta^{ab}\delta^a_i\delta^b_j(-1)^a$ and so we get now}
&=& \int \dz \delta^{ab}(a+m)_a(a+n)_{a+1} z^{m+n-1} \k^{2a+1}
(-1)^a\nonumber\\
	&=& -\k^{2a+1}m(m^2-1)\ldots(m^2-a^2)\delta^{ab}\delta_{m+n,0}
\enn
Thus the algebra (\ref{eq.d3}) is a positive-definite W-algebra since the
central term is non-degenerate. Following the theoretical framework we
laid out before, we can now restrict our attention to the algebra of
the modes of the vpa. For the vpa the central term vanishes and
the second term in (\ref{eq.d3}) is now easy to compute.
\eqq
&& \int \dz \Tr( j^W(z) [dW^{a(0)}_m, dW^{b(0)}_n] ) \nonumber\\
	&=& \int \dz \Tr( \left( \sum_c W^c(z)E^c \right) \nonumber\\
	&&\quad\times\quad \left[ \sum_{i=0}^a (a+m)_i (-\k)^i
        z^{m+a-i} E^{2a-i,a}
	, \sum_{j=0}^b (b+n)_j (-\k)^j z^{n+b-j} E^{2b-j,b} \right]
	) \k^{i+j} \nonumber\\
	&=& \sum_c W^c_{m+n} \sum_{i=0}^a \sum_{j=0}^b (a+m)_i (b+n)_j
	(-\k)^{i+j} \Tr( E^c [E^{2a-i,a},E^{2b-j,b}] ) \k^{i+j} \;,
\nonumber\\
	&=& \sum_c W^c_{m+n} \sum_{\lambda=\max(b,c)}^{\min(a+b,b+c)}
	(a+m)_{(a+b-\lambda)} (b+n)_{(\lambda-c)}
\left\{ \matrix{c&b&a\cr 0&a-b+\lambda&2b+c-\lambda}
\right\}\k^{a+b-c}
\;,
\nonumber
\enn
{\noindent remembering that the trace gives a delta function
$\delta(c+a+b,2a+2b-j-i)$.}

The third term is more complicated,
\eqq
&\!T_3&\equiv \k\int \dz  \Tr(dW^{a(0)}_m (dW^{b(1)}_n)')\nonumber\\
	&=& \int \dz
	\sum_{l=0}^a z^{a+m-l}(a+m)_l (-1)^l \nonumber\\
	&& \;\times\;
	\sum_{d,c} \sum_{i=0}^{b-1} \sum_{p=0}^{c-1} \sum_{j=1}^{c-p}
	(-1)^{p+j} \partial^j ( W^d(z)\partial^{i} z^{p+b})
	\cul d,b,c,0,2b-i,p. \Tr(E^{2a-l,a} E^{2c-j-p,c})\k^{i+j+l}
 \nonumber\\
\noalign{\noindent The last trace gives us $\delta(l,a), \delta(a,c),
\delta(j,c-p)$ and so we get now}
	&\!T_3&= \int \dz
	\sum_d   \sum_{i=0}^{b-1} \sum_{p=0}^{a-1}
	(-1)^{a} z^{m} \partial^{a-p}
	( W^d(z)\partial^{i}z^{p+b})
	\cul d,b,a,0,2b-i,p. \k^{i+2a-p}(a+m)_a
	\nonumber\\
	&=& \int \dz \sum_d \sum_{i=0}^{b-1} \sum_{p=0}^{a-1}
	(-1)^{m} (\partial^{a-p} z^{m}) (\partial^{i}z^{p+b})
	W^d(z) \cul d,b,a,0,2b-i,p. \k^{i+2a-p}(a+m)_a\nonumber\\
	&=& \int \dz \sum_d \sum_{i=0}^{b-1} \sum_{p=0}^{a-1}
	(-1)^{m} z^{m + p - a + p + b - i \blank{m+n+d-a-i+p}}  \nonumber\\
	&&\;\times\; (a+m)_a
	(m)_{(a-p)} (n+b)_{(i)}
	W^d(z) \cul d,b,a,0,2b-i,p. \k^{i+2a-n}\nonumber\\
\noalign{\noindent This last term in curly braces represents a trace,
which gives us $\delta(i, b + p - d - c)$, and so we get, with
$\lambda =2b+d-p$,}
	&\!T_3&= \! \sum_d W^d_{m+n} \!\!\sum_{\lambda=b+d+1}^{2b+d}
\!\!(a+m)_{(a+b-\lambda)} (b+n)_{(\lambda-d)} 	\cul
d,a,b,0,a-b+\lambda,2b+d-\lambda. \k^{a+b-d}\nonumber\\ 	&=&
\!\!\!\!\sum_d W^d_{m+n} \!\!\sum_{\lambda=b+d+1}^{b+d+n} \!\!
(a+m)_{(a+b-\lambda)} (b+n)_{(\lambda-d)} 	\cul
d,a,b,0,a-b+\lambda,2b+d-\lambda. \k^{a+b-d}
\;,
\label{eq.d5}
\enn
where we used the fact that for $\lambda>b+d+n$ the
second of the two Pochhammer symbols vanishes.

By similar reasoning we can evaluate the fourth term
of  (\ref{eq.d3}) and when we put them all together we obtain for the
vpa terms $\{W^a_m, |m|\le a\}$,
\eqq
\{ W^a_m, W^b_n \} &=& \sum_c W^c_{m+n}
	\sum_{\lambda=max(c,b-m)}^{min(b+c+n,a+b)}
	(a+m)_{(a+b-\lambda)} (b+n)_{(\lambda-c)} \nonumber\\
	&&\quad\times\quad
	\cul c,a,b,0,a-b+\lambda,2b+c-\lambda. \k^{a+b-c}\quad + O(W^2) \label{eq.d6}
\enn
This establishes the linearised vpa commutation relations.
We can now compare them with the commutation relations of $g$ in a
particular basis.
We know that $W[I_-]_{\pm1,0}$ form the $sl(2)$ embedded inside $g$, and so
it is convenient for us to take the modes of the (quasi-)primary
fields $W^a_m$ to correspond to the representation of $sl(2)$ with
highest weight $E^a$. If we take the correlation to be
\eq
W^a_m \cong E^{a-m,a} f(a,m)
\en
where $f(a,m)$ are some constants, then the fact that $W[I_-]_{\pm 1,0}$
are the embedded $sl(2)$ tells us that $f(a,m)$ is given by
\eq
f(a,m) = \mu(a) (-\k\lambda)^{a+m} (a+m)!  / (2a)! \;,
\en
where $\lambda$ is defined in equation (\ref{eq.ldef}).
We may now evaluate the commutator of two of these elements of $g$,
and we find that we recover the commutation relations (\ref{eq.d6})
exactly for $\mu(a) = (2a)! $, and so
\eq
W^a_m \cong E^{a-m,a} (-\k\lambda)^{a+m} (a+m)!
\en
completes the identification of the vacuum preserving modes of the
W-algebra with the algebra $g$ itself, up to quadratic terms in the
fields $W$. Moreover the modes $L_1,L_0,L_{-1}$ are clearly associated
to $I_+,I_0,I_-$. If one examines the commutation relations more
carefully and normalises the algebra correctly according to
(\ref{eq.stqumodes2}), one sees that indeed the couplings $f^{ab}_c$
are $O(1)$ in the central charge, thus bearing out our expectations
for a classical W-algebra.

This clarifies the relationship between the linearised vpa and the
generalised Drinfeld-Sokolov reduction. It follows that there exists
at least
one classical W-algebra for every $g$ and every integral $su(1,1)$
embedding. If this W-algebra is unique, then the Drinfeld-Sokolov
reductions completely saturate the possibilities for bosonic integrally
weighted W-algebras.
We return to this point in the conclusion.

\section{Lie algebras and $sl(2)$ embeddings}

\def\topspace{\vphantom{\vrule height 3ex depth 0pt}}
\def\bottomspace{\vphantom{\vrule height 0pt depth 2ex}}

In this section, we shall use the theory of finite Lie algebras and
their three-dimensional subalgebras together with what we have learnt
in the preceding sections to look at various aspects of reductive W-algebras.
We briefly comment on a possible relation between automorphisms of the
linearised vpa and homomorphisms of the W-algebra. Then we give a complete
list of positive-definite W-algebras without Kac-Moody components by
classifying
all su(2) embeddings of semi-simple Lie algebras with trivial center.
We enumerate the algebras which, in addition, contain no generating
spin-2 fields besides the Virasoro algebra.

First, let us consider the case where the $su(1,1)$ is not a maximal
subalgebra of $g$;
that is there exists some algebra $h$ such that $su(2)\subset h \subset g$.
In the limit that $c\to \infty$ we expect that some subalgebra of generating
fields which includes the Virasoro algebra closes. This, however, does
{\em not} imply
that these generating fields generate some $W$-algebra associated
with $su(2)\subset h$,
 which is a subalgebra of the $W$-algebra associated with
$su(2)\subset g$.
As an example we can consider the algebra $WA_3$
associated with the principal $su(2)$ embedding in $A_3$ which is generated
by fields of wieght $2,3,$ and $4$. However we can
write $su(2)\subset B_2\subset A_3$ in this case,
where the spin $2,4$ fields are
associated with the first embedding. An inspection of the operator product
expansion of the spin $4$ field with itself reveals that it contains a
term which is the composite field associated with the square of the spin $3$
field. The coupling to this term vanishes in the limit that $c\to \infty$, but
the spin $2$, $4$ algebra does not close on itself for finite central charge,
and is distinct from $WB_2$ which we would associate with the first embedding.
However, we suspect that a weaker statement is true. If
$\tau$ is an automorphism
of $g$ for which $h$ is the stable subalgebra, this gives
an automorphism of the vacuum preserving modes in the limit that
$c\to \infty$. Since $\tau(L)=L$ it
defines some homomorphism which maps a generating primary field
to some linear
combination of primary fields of the same weight. We shall assume
that this can be used to define a homomorphism on the Verma module of the
W-algebra associated with $g$. The subalgebra which is stable under
this homomorphism will contain only those generating fields associated with
$h$, but will in general require additional generators which will be
composites in all the generating fields of the algebra. In the example above,
if we choose the simple roots of $A_3$ to be $e_1-e_2,e_2-e_3,e_3-e_4$ then
the automorphism which preserves its $B_2$ subalgebra is given by
\begin{eqnarray*}
e_1 & \to &  -e_4\\
e_2 & \to &  -e_3
\end{eqnarray*}
and the induced homorphism on the $WA_3$ algebra is simply
spin-$3 \to-$spin-3. The subalgebra which is stable under this homomorphism
is the smallest closed algebra containing the spin $2,4$ fields.
If it is the case that the resulting W-algebra does not have a good classical
limit, which we know to be true in the case that we reduce a super W-algebra
in this way, then this sort of consideration may provide a powerful tool
for constructing deformable algebras which do not have a good classical limit.

As a second topic of interest, we shall now give a complete list of
the possibilities for bosonic W-algebras with good classical limit
which have no Kac-Moody components.
If we want there to be no spin 1 fields in the
$W$-algebra we require that there
are no singlets in the decomposition of the adjoint representation of $g$
under $su(1,1)$. Alternatively, if we decompose $g$ w.r.t. the $I_0$
member of the preferred $sl(2)$, we may express this by saying that
${\rm dim } g_1={\rm dim }g_0$,
or by saying that the centraliser of $su(1,1)$ in $g$ is zero.
We classify all the examples of $su(1,1)$ embeddings where this
is so.

First we need to borrow some notation from Dynkin \cite{Dynk}.
For ease of exposition
we shall revert to the real compact form of the Lie algebras, since the
form of the algebra is not important for the argument.
A regular subalgebra of $g$
is a subalgebra whose root system is simply a subset of the root system of $g$.
A subalgebra of $g$ is called an $R$-subalgebra if it is contained in some
proper regular subalgebra of $g$. Otherwise it is called an $S$-subalgebra.
$S$-subalgebras have the properties that

{\obeylines
{(i)} they are integral
{(ii)} ${\rm dim }g_1={\rm dim }g_0$. }

\noindent
Thus the condition that an $su(2)$ subalgebra by an $S$-subalgebra is
sufficient
for producing a $W$-algebra without Kac-Moody components, but as it
turns out it is not necessary. In the other cases there must exist
some proper regular subalgebra of $g$ which contains $su(2)$ and
moreover its rank must be equal to that of $g$, since otherwise it is
easy to prove that some memeber of the  Cartan subalgebra will commute
with it. The classification of all $su(2)$ subalgebras of the
exceptional Lie algebras whose centraliser vanishes has  been given
in \cite{Dynk}. We reproduce these results in Table 1.
The algebra $g$ is given in the first column. The second column
gives the characteristic specifying the $su(2)$ embedding while
the third summarises the weights (with degeneracies in parentheses)
of
the generating fields of the corresponding $W$-algebra.
The next column gives the minimal subalgebra(s) of $g$ which
contain the $su(2)$. A $P$ in the final column indicates that the
embedding is principal.
We now deal with the remaining classical examples.

\vskip 7mm
\noindent
(i) $su(n)$

\vskip 3mm
\noindent
The only $su(2)$ $S$-subalgebra of $A_n$ is given by the principal embedding.
Any other candidate is a subalgebra of one of the maximal regular subalgebras
of $su(n)$ and hence we can write
$su(2)\subset su(p)\oplus su(q)\oplus u(1)\subset su(n)$.
Furthermore it is clear that $su(2)\subset su(p)\oplus su(q)$ so that
the $u(1)$ factor commutes with it. Thus no other algebra has zero
centraliser.

\vskip 7mm
\noindent
(ii) $sp(n)$

\vskip 3mm
\noindent
Again the only $su(2)$ S-subalgebra of $sp(n)$ is the principal subalgebra.
For the other subalgebras we may write
\eq
su(2)\subset c(p_1)\oplus c(p_2)\subset c(n) \;\;\;\; p_1+p_2=n
\en
If the copy of the $su(2)$ is not principal in one of the $c(p_i)$ we can
decompose this further till we have
\eq
su(2)\subset c(p_1)\oplus .....\oplus c(p_i)\subset c(n) \;\;\;\; \sum_j p_j=n
\en
where the copies of $su(2)$ in each $c(p_i)$ are principal. We can decompose
the adjoint representation of $c(n)$ with respect to this direct
sum of $c(p_i)$ within which the $su(2)$ are principal.
We find that
\begin{eqnarray*}
{\rm adj }c(n) =&& ({\rm adj}c(p_1)\otimes 1 \otimes \ldots \otimes 1)
\oplus...\oplus (1 \otimes \ldots \otimes 1 \otimes {\rm adj } c(p_i))
\\
& \bigoplus_{j,k}  &
  1 \otimes \ldots 1 \otimes 2p_j \otimes 1 \ldots 1 \otimes 2p_k
\otimes 1 \ldots \otimes 1 \;,
\end{eqnarray*}
where we have denoted the $2p$ dimensional representation of $c(p_i)$
by $2p_i$.
The adjoint representation of $c(p_i)$ contains no singlets when decomposed
with respect to a principal $su(2)$. The $2p_i$ representations deomposes into
a single irreducible representation of this $su(2)$ and so the tensor
product
$...2p_i\otimes ...2p_j..$ decomposes with respect to the diagonal su(2) to
give $|2p_i-2p_j|\oplus...\oplus 2p_i+2p_j$.
{}From this we can see that adj$c(n)$ contains a trivial representation
of the diagonal $su(2)$ subalgebra if and
only if $p_j=p_k$ for some $j\neq k$.

\vskip 7mm
\noindent
(iii) $so(n)$

\vskip 3mm
\noindent
The argument in this case is a little more involved. $b_n$ again possesses no
other $su(2)$ $S$-subalgebras other than its principal. $d_n$ possesses
${\rm int}[(n-2)/2]$ $S$-subalgebras, but in fact none of these are maximal
(not even the principal) and they correspond to the embeddings
$su(2)\subset so(p)\oplus so(q)\subset so(2n)$ where $p+q=2n$ and
$p,q$ are both odd.
For our purposes it will be convenient to think of $so(4)$ as simple. Its
principal subalgebra is maximal.

Now starting with any $su(2)$ embedding in $so(n)$ it is straightforward to
show that
\eq
su(2)\subset so(p_1)\oplus so(p_2)\oplus....\oplus so(p_i)\subset so(n)
\en
where $p_i=3,4,5,7,9,....$ and the copy of $su(2)$ in each simple ideal
is principal.
For $\bigoplus_j so(p_j)$ to be maximal in $so(n)$
and hence to have trivial centraliser we need that \hbox{$n-\sum_j p_j
>2$}. Again we can decompose
the adjoint representation of $so(n)$ with respect to this direct product of
principal $su(2)$ subalgebras and we find that
\begin{eqnarray*}
{\rm adj }so(n)  &=   & \bigoplus_i 1 \otimes 1 \otimes \ldots \otimes
{\rm adj }so(p_i) \otimes 1 \otimes \ldots \otimes 1
 \\
 & \bigoplus_{j,k}&
  1 \otimes \ldots 1 \otimes p_j \otimes 1 \ldots 1 \otimes p_k
\otimes 1 \ldots \otimes 1 \;,
\end{eqnarray*}
where we have denoted the $p_j$ dimensional representation of
$so(p_j)$ by $p_j$.
Again this will have a trivial representation with respect to the
diagonal $su(2)$ if and only if $p_j=p_k$ for some $j\neq k$. Notice that
if one of the $p_j=4$, the associated embedding is non-integral. This is the
only example of a W-algebra with no Kac-Moody components which is not
positive-definite and can be obtained in this way. The results
for $su(2)$ embeddings in classical algebras are summarised in Table 2.

It is also straightforward to extend the above analysis to search for
W-algebras with no Kac-Moody components and no
other spin-2 fields than that associated with the
Virasoro algebra. If there exists some algebra $h$ such that
$su(2)\subset h\subset g$, $h$ is not simple, and $su(2)$ is
embedded diagonally in more than one of the simple ideals of $h$
then there exists more than one spin 2 field. This is because the
decomposition of the direct product of $N$ copies of $su(2)$ with
respect to its diagonal subalgebra contains $N$ spin-$1$
representations. The cases where no such $h$ exist are easy to read
off from the tables, and are marked with a check-mark in Table 1. For
the classical algebras, it is clear that only the principal embeddings
result in only one spin-2 field.

\vfill\eject

\def \threeonearrow {$\equiv$\hskip -3.5mm $>$}
\def \twoonearrow  {$=$\hskip -3.5mm $>$}

\begin{center}
\begin{tabular}{|c|c|ccccccc|l|c|c|}
\hline \topspace
$G$&$Index$& \multicolumn{6}{c}{\hskip .5cm Characteristic}&&
\hskip 1.9cm Spins & H& spin-2	\\
\hline \topspace
  $E_8$	& 40&0&0&0&1&0&0&0&2(10),3(10),4(10),5(6),6(4)& $A_4\oplus A_4$	&\\
&&&&&&0&&&&& \bottomspace \\
  $E_8$	& 88&1&0&0&0&1&0&0&2(4),3(4),4(5),5(3),6(6)&$E_6\oplus A_2$& \\
&&&&&&0&&&7(2),8(3),9& $B_4\oplus B_3$&\bottomspace \\
  $E_8$	& 120&0&1&0&0&1&0&0&2(3),3,4(5),5(3),6(3)& $B_5\oplus B_2$ &\\
&&&&&&0&&&7(3),8(3),9,10(2)&&\bottomspace \\
  $E_8$	& 160&1&1&0&0&1&0&0&2(4),3,4(2),5(3),6(3)& $E_7\oplus A_1$& \\
&&&&&&0&&&8(3),9(3),10(2),12&&\bottomspace \\
  $E_8$	& 184&0&1&0&0&1&0&1&2(3),3,4,5,6(4),7(2)& $B_6\oplus A_1$& \\
&&&&&&0&&&8(3),9,10,11,12(2)&&\bottomspace  \\
  $E_8$	& 232&1&1&0&0&1&0&1&2(2),3,4(2),6(3),7& $E_7\oplus A_1$& \\
&&&&&&0&&&8(2),9(2),10,11,12(2),14&&\bottomspace  \\
  $E_8$	& 280&1&0&1&0&1&0&1&2,3,4,5,6(2),8(3)& $B_7$& \checkmark \\
&&&&&&0&&&9,10(2),12(2),14,15&&\bottomspace  \\
  $E_8$	& 400&1&1&1&0&1&0&1&2(2),5,6(2),8,9,10(2)& $E_7\oplus A_1$& \\
&&&&&&0&&&12,14(2),15,18&&\bottomspace  \\
  $E_8$	& 520&1&1&0&1&0&1&1&2,4,6,8,9,10,12(2)& $E_8$& \checkmark \\
&&&&&&1&&&14,15,18,20&&\bottomspace \\
  $E_8$	& 760&1&1&1&1&0&1&1&2,6,8,10,12,14,15,18& $E_8$&\checkmark  \\
&&&&&&1&&&20,24&&\bottomspace \\
  $E_8$	&1240&1&1&1&1&1&1&1&2,8,12,14,17,20,24,30& $E_8$&\checkmark P  \\
&&&&&&1&&&&&\bottomspace \\
  $E_7$	&  39&&1&0&0&1&0&0&2(6),3(4),4(5),5(3),6(3)& $A_5\oplus A_2$ &\\
&&&&&&0&&&&$D_6\oplus A_1$&\bottomspace\\
  $E_7$	&  63&&1&0&0&1&0&1&2(4),3(2),4(3),5(2),6(4)& $D_6\oplus A_1$ &\\
&&&&&&0&&&7,8&&\bottomspace\\
  $E_7$	& 111&&1&1&0&1&0&1&2(2),3,4(2),5,6(3)& $D_6\oplus A_1$ &\\
&&&&&&0&&&8(2),9,10&&\bottomspace \\
  $E_7$	& 159&&1&0&1&0&1&1&2(2),,4,5,6(2),8(2)& $F_4\oplus A_1$ &\\
&&&&&&1&&&9,10,12&&\bottomspace \\
\hline\noalign{\medskip}
\multicolumn{12}{c}{\hbox{Table 1}}
\end{tabular}
\end{center}

\vfill\eject

\begin{center}
\begin{tabular}{|c|c|ccccccc|l|c|c|}
\hline \topspace
$G$&$Index$& \multicolumn{6}{c}{\hskip .2cm Characteristic}&&
\hskip 1.5cm Spins & H & spin-2	\\
\hline \topspace
  $E_7$	& 231&&1&1&1&0&1&1&2,4,6(2),8,9,10,& $E_7$ &\checkmark \\
&&&&&&1&&&12,14&&\bottomspace\\
  $E_7$	& 399&&1&1&1&1&1&1&2,6,8,10,12,14,18& $E_7$&\checkmark P \\
&&&&&&1&&&&&\bottomspace\\
  $E_6$	& 36&&&1&0&1&0&1&2(3),3(3),4(2),5(2),6(2)& $A_5\oplus A_1$	&\\
&&&&&&0&&&& &\bottomspace\\
  $E_6$	& 84&&&1&1&0&1&1&2,3,4,5,6(2),8,9&$C_4,G_2$ &\checkmark \\
&&&&&&1&&&& &\bottomspace\\
  $E_6$	&156&&&1&1&1&1&1&2,5,6,8,9,12& $F_4$ &\checkmark P\\
&&&&&&1&&&&&\bottomspace\\
  $F_4$	&  12&&&0&1&\twoonearrow &0&0&2(6),3(4),4(2)& $A_3\oplus A_1$
&\bottomspace\\
  $F_4$	&  36&&&0&1&\twoonearrow &0&1&2(3),3,4,5,6(2)& $C_3\oplus A_1$
&\bottomspace\\
  $F_4$	&  60&&&1&1&\twoonearrow &0&1&2,3,4,6(2),8& $B_4$
&\checkmark \bottomspace\\
  $F_4$	& 156&&&1&1&\twoonearrow &1&1&2,6,8,12& $F_4$
&\checkmark P\bottomspace\\
  $G_2$	& 4&&&&&1&\threeonearrow&0 &2(3),3& $A_1\oplus A_1$
&\bottomspace\\
  $G_2$	& 28&&&&&1&\threeonearrow &1&2,6& $G_2$
&\checkmark P\bottomspace\\
\hline\noalign{\medskip}
\multicolumn{12}{c}{\hbox{Table 1 (cont)}}
\end{tabular}
\end{center}
\vskip.5cm

\begin{center}
\begin{tabular}{|c|c|l|c|}
\hline \topspace
$G$&$Index$&
Spins & H \\
\hline \topspace
  $A_n$	& ${n(n+1)(n+2)\over 6}$ &2,3,4,...,n+1& $A_n$
\bottomspace \\
  $B_n$	& ${n(n+1)(2n+1)\over 3}$ &2,4,6,...,2n& $B_n$
\bottomspace \\
  $C_n$	& ${n(2n+1)(2n-1)\over 3}$ &2,4,6,...,2n& $C_n$
\bottomspace \\
  $D_n$	& ${n(n-1)(2n-1)\over 3}$ &2,4,6,...,2n-2,n& $B_n$
\bottomspace \\
  $C_n$	& $\sum_i{n_i(2n_i+1)(2n_i-1)\over 3}$ &$\sum_i 2,..,2n_i+$
& $\bigoplus_i C_{n_i}$   \\
&&$\sum_{i>j} |n_i-n_j|+1,...,n_i+n_j+1$
&$ \sum_i n_i=n,n_i\neq n_j$\bottomspace \\
  $so(N)$& $\sum_i {n_i(n_i-1)(n_i+1)\over 12}$ &$\sum_i 2,..,{n_i-1\over 2}+
 2\sum_{i:n_i=4}2+$&$\bigoplus_i so(n_i)$\\
  &$+\sum_{i:n_i=4}2$ & $\sum_{i>j} |n_i-n_j|+1,...,n_i+n_j+1$
&$\sum_i n_i=n,n-1,n_i\neq n_j $ \\
&&&$n_i\in 3,4,5,7,.$ \bottomspace \\
\hline\noalign{\medskip}
\multicolumn{4}{c}{\hbox{Table 2 }}
\end{tabular}
\end{center}
\vskip1cm

\section{Conclusions}

In this paper we have found a connection between a general class
of W-algebras and finite Lie algebras. A crucial role in our arguments
was played by the {\em vacuum-preserving-algebra}(vpa)
which is the closed subalgebra
of modes which annihilate both right and left vacuua.
For `linear' W-algebras one finds that the vpa contains a finite subalgebra
which provides a useful tool for studying the
properties of theories invariant under these W-algebras.  To extend
this idea to more general non-linear quantum W-algebras, it
became necessary to consider {\em deformable} W-algebras which are
defined for a range
of $c$ values in the classical limit $c\to\infty$, and the subclass
of algebras which behave `well' under this limit.
A natural criterion which arises is that of positive-definitness of
the W-algebra, which essentially ensures that all the fields are
important to the structure of the algebra. Reductive algebras are
algebras which
have positive-definite classical limits.
As a result we were
able to assign to each reductive
W-algebra a finite Lie
(super-)algebra and an embedding of $su(1,1)$.
The field content of the W-algebra is encoded in this embedding, with
each representation of $su(1,1)$ in the decomposition of the Lie
algebra being associated to one of the Virasoro primary fields; the
weight of that field being equal to (1 + the dimension)/2.
By considering the structure of the commutation relations of the
W-algebra, combined with the Virasoro Ward identities, we were also
able to
show that this finite algebra was restricted to be of the form of a
direct sum of a semi-simple algebra and an abelian
algebra\reductive{, namely a \compact\  Lie algebra.}{.}
This condition places considerable restrictions on the possible field
contents of W-algebras and on their commutation relations.
As an example of the ideas presented, we considered the classical
Poisson bracket algebras of generalised Drinfeld-Sokolov type. The
analysis here held out our theoretical predictions -- to each such
W-algebra we were able to assign a finite Lie algebra and an $su(1,1)$
embedding in that algebra. Conversely we used the construction to demonstrate
the existence of a classical W-algebra associated with each such pair.

Our work suggests that W-algebras can be divided broadly into three categories:
the reductive algebras considered in this paper, other deformable algebras, and
`non-deformable' algebras which are only associative for specific values of
$c$.
There are several questions which present themselves concerning each category.

Firstly, although we have demonstrated the existence of a classical W=algebra
associated with each finite algebra and $su(1,1)$ embedding, it is not clear
as yet whether one can
actually find a quantum W-algebra for each such embedding.
The quantisation of these
models has a lengthy history and is by no means over yet
\cite{BGer3,FLuk2,KWat2,FFre3}, although there seem to be good arguments
in favour of their existence.
If one can find quantum W-algebras which satisfy the conditions of
section 2, namely having a good classical limit which is  \posdef ,
then the question obviously arises, are they unique? That is, to each
such embedding can one uniquely ascribe a quantum W-algebra? We know
of no counterexamples.

Our conditions, although they catch many of the W-algebras which have been
studied to date which have proven useful in conformal field theory,
still exempt many W-algebras. Indeed we present such an exception with
fields of spins 2,4, and 6. In section 5 we have presented what we
hope will be a useful approach to the study of these algebras, namely
automorphisms of Lie algebras which preserve a subalgebra. For the
case we presented this was a $\Bbb Z_2$ automorphism which preserved
the $B_2$ subalgebra of $A_3$. We hope that we can extend
this to other cases. Certainly the idea of dividing out by a finite
group action is not new, but rather the idea that we may be able to
ascribe a W-algebra uniquely to each such action, and even reconstruct
the larger algebra from the smaller, is. There are good reasons to believe
that the absence of a good classical limit implies strong constraints
on the unitary representations of W-algebras, and we hope to return to
this and other topics in the future.

As even more distant projects we can mention the idea that one may be
able to show that each W-algebra which occurs for a specific set of
$c$-values is simply the extension of a {\it deformable} W-algebra by
primary fields of integer spin. Thus, it may be instructive to look
for `maximal' deformable subalgebras of such W-algebras.

\blank{
for the sake of the automatic refereeing, maybe we should say Witten,
Witten, Witten, black holes, black holes, and maybe ginsparg
}

\vskip 1cm

\noindent
PB would like to thank the DOE of the U.S.A.for support under grant
number DEFG02-90-ER-40560 and the NSF of the U.S.A. under grant
PHY900036, and GMTW would like to thank the SERC of the U.K. for a
research assistantship. GMTW would like to thank the EFI for
hospitality at the initiation of this work. PB and GMTW would like to
thank the Institute of Theoretical Physics at the University of
California, Santa Barbara, for hospitality, and where they were in
part supported by the National Science Foundation under Grant
No.~PHY89-04035. GMTW would like to thank Trinity College Cambridge
for a Rouse-Ball travelling studentship.

We gladly thank H.~G.~Kausch for many useful conversations during the
course of this project.
GMTW would like to thank C.~F.~Yastremiz for useful comments on the
manuscript.

\blank{
CITATIONS TO GO IN
\cite{BGer2,BGer3,BGer4,GERV,FRRT1,ORTW1,OWip1,BFFOW2,BTDr1,Fehe1,Bers1,BOog1,PRIN,AS}
}

\end{document}